\documentclass{aa}

\usepackage[varg]{txfonts}
\usepackage{natbib}
\usepackage{graphicx}

\bibpunct{(}{)}{;}{a}{}{,} % to follow the A&A style

\makeatletter
\renewcommand*\aa@pageof{, page \thepage{} of \pageref*{LastPage}}
\makeatother

\begin{document}

    \title{Bar-spiral interaction induces radial migration and star formation bursts}

    \author{L. Marques \inst{\ref{aip},\ref{uniP},\ref{mauca}},
            I. Minchev \inst{\ref{aip}},
            B. Ratcliffe \inst{\ref{aip}},
            S. Khoperskov \inst{\ref{aip}},
            M. Steinmetz \inst{\ref{aip},\ref{uniP}},
            T. V. Wenger \inst{\ref{trey}},
            T. Buck \inst{\ref{tb1}, \ref{tb2}},
            M. Martig \inst{\ref{mm}},
            G. Kordopatis \inst{\ref{oca}},
            M. Schultheis \inst{\ref{oca}},
            D. B. Zucker \inst{\ref{UMq1},\ref{UMq2}}
            }

    \institute{Leibniz-Instit\"ut f\"ur Astrophysik Potsdam (AIP), An der Sternwarte 16, D-14482, Potsdam, Germany\label{aip} \\
    \email{lmarques@aip.de, iminchev@aip.de}
    \and Universit\"at Potsdam, Institut f\"ur Physik und Astronomie, Karl-Liebknecht-Str. 24-25, 14476 Potsdam, Deutschland\label{uniP}
    \and MAUCA -- Master track in Astrophysics, Universit\'e C\^ote d'Azur, Observatoire de la C\^ote d'Azur, Parc Valrose, 06100 Nice, France\label{mauca}
    \and NSF Astronomy \& Astrophysics Postdoctoral Fellow, Department of Astronomy, University of Wisconsin–Madison, Madison, WI 53706, USA\label{trey}
    \and Universit{\"a}t Heidelberg, Interdisziplin{\"a}res Zentrum f{\"u}r Wissenschaftliches Rechnen, Im Neuenheimer Feld 205, D-69120 Heidelberg, Germany\label{tb1}
    \and Universit{\"a}t Heidelberg, Zentrum f{\"u}r Astronomie, Institut f{\"u}r Theoretische Astrophysik, Albert-Ueberle-Straße 2, D-69120 Heidelberg, Germany\label{tb2}
    \and Astrophysics Research Institute, Liverpool John Moores University, 146 Brownlow Hill, Liverpool L3 5RF, UK\label{mm}
    \and Universit\'e C\^ote d’Azur, Observatoire de la C\^ote d’Azur, CNRS, Laboratoire Lagrange, 06000 Nice, France\label{oca}
    \and School of Mathematical and Physical Sciences, Macquarie University, Sydney, NSW 2109, Australia\label{UMq1}
    \and Macquarie University Research Centre for Astrophysics and Space Technologies, Sydney, NSW 2109, Australia\label{UMq2}
    }

    \date{}

    \abstract {Central bars and spirals are known to strongly impact the evolution of their host galaxies, both in terms of dynamics and star formation. Their typically different pattern speeds cause them to regularly overlap, which induces fluctuations in bar parameters.}
    {In this paper, we analyze both numerical simulations of disk galaxies and observational data to study the effect of bar-spiral physical overlap on stellar radial migration and star formation in the bar vicinity, as a function of time and galactic azimuth.}
    {We study three different numerical models, two of which are in a cosmological context, as well as APOGEE DR17 data and the WISE catalog of Galactic HII regions.}
    {We find that periodic boosts in stellar radial migration occur when the bar and spiral structure overlap.
    This mechanism causes net inward migration along the bar leading side, while stars along the bar trailing side and minor axis are shifted outward. The signature of bar-spiral induced migration is seen between the bar's inner Lindbald resonance and well outside its corotation, beyond which other drivers take over.
    We also find that, in agreement with simulations, APOGEE DR17 stars born at the bar vicinity (mostly metal-rich) can migrate out to the solar radius while remaining on cold orbits.
    For the Milky Way, 13\% of stars in the solar vicinity were born inside the bar, compared to 5-20\% in the simulations.
    Bar-spiral reconnections also result in periodic starbursts at the bar ends with an enhancement of up to a factor of 4, depending on the strength of the spiral structure. Similarly to the migration bursts, these do not always happen simultaneously at the two sides of the bar, hinting at the importance of odd spiral modes. Data from the WISE catalog suggest this phenomenon is also relevant in our own Galaxy.
    }
    {}


    \authorrunning{Marques, Minchev et al.}
    \maketitle

    \nolinenumbers
    
    \section{Introduction}
    \label{section:intro}
    \paragraph{} Spiral structures are over-dense spiral-shaped regions emerging from the center of many disk galaxies.
    Their nature is still debated, between long-lived density waves with a constant rigid-body pattern speed \citep{lin1964, kalnajs1973, bertin1996}, the overlap of multiple spiral modes \citep{tagger1987,sygnet1988, quillen2011, minchev2012a}, transient material arms which would have the same rotation curve as the stars and gas \citep{grand2012a,grand2012b,baba2013} \citep[although strong bars could increase the spirals' rotation speed, see, e.g.,][]{grand2012b, roca-fabrega2013}, or temporary responses to mass clumps in the disk \citep{toomre1991,d'onghia2013}.
    Whichever their nature, spirals are very frequent in our local Universe, and 60 to 70\% of spiral galaxies, including the Milky Way (MW), also host a bar feature in their center (\citealp{eskridge2000, menendez-delmestre2007, marinova2007, sheth2008, erwin2018}; and \citealp{peters1975, binney1991, blitz1991, nakada1991, weiland1994, stanek1997} for the MW).
    %Bars are elongated non-axisymmetric structures composed of gas, stars and dust and found at the center of galaxies.
    Both the bar and the spiral structure have essential roles in the evolution of their galaxy; two relevant phenomena studied in this paper are radial migration and star formation.

    Radial migration is the change of stellar orbital angular momentum, and thus of guiding radius, and has been seen to occur in numerous N-body simulations of isolated galactic disks \citep{SB02,Roskar2008_migration,minchev2010, grand2012b}, as well as in galaxies simulated in the cosmological context \citep{minchev2013, vincenzo2020,agertz2021_vintergatan, lu2022b, okalidis2022_auriga-migr,boecker2023_tng50-migr}. Several mechanisms can explain migration. Transient spiral structures can cause large scale migration near their corotation radius as was shown by \citet{SB02, grand2012b, kawata2014_CRspiral}. The transient property here is key, because a long-lived spiral would make stars' angular momentum go back to their initial values each time the star meets a spiral arm again, instead of truly redistributing stellar angular momenta. The bar, being a more stable pattern, produces migration using different mechanisms. \citet{minchev2006, minchev2011_res-overlap} showed that the overlap of resonances of multiple patterns (such as multiple spiral modes or bar and spiral, respectively) could induce efficient migration in the disk, without requiring a transient spiral structure. This mechanism can also significantly heat stars kinematically and explain the age-velocity dispersion relation in the MW. A bar slowing down rapidly during and right after its formation can also sweep out kinematically cold stars trapped in its resonances, without kinematically heating them, as shown by \citet{Khoperskov2020_barres-sweep}. Apart from these secular processes, mergers can similarly provoke radial mixing by introducing an abrupt change in potential \citep[e.g.,][]{quillen2009_mergermigr, bird2012}.
    By moving stars away from their birth place, this process can explain the scatter in the age-metallicity relation in the solar vicinity \citep[e.g.,][]{Roskar2008_migration,schoenrich2009_chem-migr} by flattening the radial metallicity gradient at the same time \citep{minchev2013}. Indeed, radial migration has been seen in simulations to happen mostly outward \citep[e.g.,][]{khoperskov2021, agertz2021_vintergatan, vincenzo2020, renaud2024} (although this is likely due to the stellar density decreasing outward) so that metal-rich stars reaching the solar neighborhood (SNd) from the inner disk would increase the local mean metallicity, thus flattening its radial gradient at larger radius. Since migrators reaching the SNd are mostly indistinguishable from locally born stars in their orbital properties, chemical abundances and age information are needed to identify their birth locations, using the technique of chemical tagging \citep{freeman2002_chemtag}. This is how the first observational evidence for radial migration was hinted by \citet{grenon1972,chiappini2009obs-migr,boeche2013}, and later explicitly found by \citet{kordo2015_obs-radmigr} using the fact that the local star formation history cannot explain the metallicity of the super-metal rich ([Fe/H] > 0.25 dex) stars observed in the SNd \citep{chiappini2003_chemevol, spitoni2019}. In parallel, it is essential to understand how much radial migration must have occurred given the presence of a bar and spiral arms in the MW. Recently, a method to derive birth radii with virtually no dependence on any assumptions on the chemical evolution of the MW has been developed by \citet{minchev2018_rb, lu2022b, ratcliffe2024c}, and already used by \citet{ratcliffe2023_rb, ratcliffe2024c} on APOGEE DR17 red giants to study the evolution of metallicity gradients. This has huge potential to help us quantify the radial dependence of migration in the MW. However, only simulations for now can give us the time evolution of migration.

    The relation between bars or spirals and star formation (SF) is also widely researched, and conclusions are still debated. SF is triggered inside molecular clouds when the equilibrium between pressure and gravity is broken in favor of gravity. Overdense regions like spiral arms are thus natural locations for SF to take place. Young stars and molecular clouds are indeed predominantly found in spiral arms \citep{levine2006,poggio2021,castroginard2021,gaiacollab2023}. The chemical signature of spiral arms also shows higher metallicity in the spiral arms compared to inter-arm regions \citep{poggio2022, hackshaw2024, barbillon2025}. It is however unclear whether spiral arms act as gatherers or triggers of SF. The way to distinguish between these two possibilities is by comparing star formation efficiencies (SFE) between arm and inter-arm regions. Observations are still inconsistent and contradictory. In theory, the gas is shocked when it enters the dense spiral arms, which should trigger SF, as predicted by density wave theory. Several authors found evidence that SFE indeed seemed higher in the spiral arms, thus encouraging the triggering scenario  \citep{lord1987, vogel1988, cepa1990, lord1990, knapen1996, seigar2002, cedres2013, karapetyan2018}. However, other authors found that SFE was overall similar in spiral arms and interarm regions, despite local fluctuations \citep{elmegreen1986, foyle2010, moore2012, rebolledo2012, eden2015,ragan2018, querejeta2021, querejeta2024, sun2024}. The way in which the bar influences SF varies with bar parameters \citep{geron2021,geron2023,geron2024}, and largely depends on temporal and spatial scales. Barred galaxies are believed to funnel gas to the central regions \citep{combes1988, combes1994, sakamoto1999, carles2016} thus fuelling the higher star formation rates (SFR) there \citep{sakamoto1999, alonso-herrero2001, ellison2011, coelho2011, lin2020}. However, SF is often quenched along the bar \citep{reynaud1998, sakamoto1999, zurita2004, khoperskov2018, diaz-garcia2020, fraser-mckelvie2020}, and enhanced again at the bar ends \citep{reynaud1998, sakamoto1999, watanabe2011, diaz-garcia2020, fraser-mckelvie2020, maeda2020, geron2024}, right at the interface with spiral arms. 

    \citet{hilmi2020} showed, using two different cosmological simulations, that the parameters of the bar, namely strength, length and pattern speed, all fluctuate on the same timescale as a result of the bar's interaction with the multiple spiral modes. Indeed, each time the bar catches up with a spiral arm (spirals are usually slower than the bar), they will overlap for a short amount of time which can bias bar parameters' measurements (e.g., the bar can appear longer if connected to a spiral arm). \citet{vislosky2024} studied how this affects the butterfly pattern seen in the Gaia DR3 radial velocity field and used it to constrain the current state of the Milky Way.
    
    In this paper, we investigate whether the bar-spiral regular overlap also impacts radial migration and SF, especially at the interface region. Indeed, it is expected that each overlap creates a transitory overdensity at the bar-spiral interface, which could redistribute angular momentum similar to transient spirals, while also producing bursts of SF. For the first time, we study how migration strength varies in the vicinity of the bar with both azimuthal angle and time. We use one N-body simulation of an isolated galaxy, as well as two cosmological simulations, and compare their SNd-like regions in the last snapshots with the birth radius distributions derived in \citet{ratcliffe2024c} for an APOGEE DR17 sample of red giants.

    Sections \ref{sec:sim} and \ref{sec:apogee} present the simulations and the APOGEE sample used. The results are presented and discussed in Sec. \ref{sec:radmigr} and \ref{sec:starformation} where we expand on the time evolution of the migration strength near the bar ends, the impacts for the SNd and the time evolution of the SFR at the bar ends. We finally conclude in Sec. \ref{sec:conclu}.

 \begin{figure*}
    \centering
        \includegraphics[width=18cm]{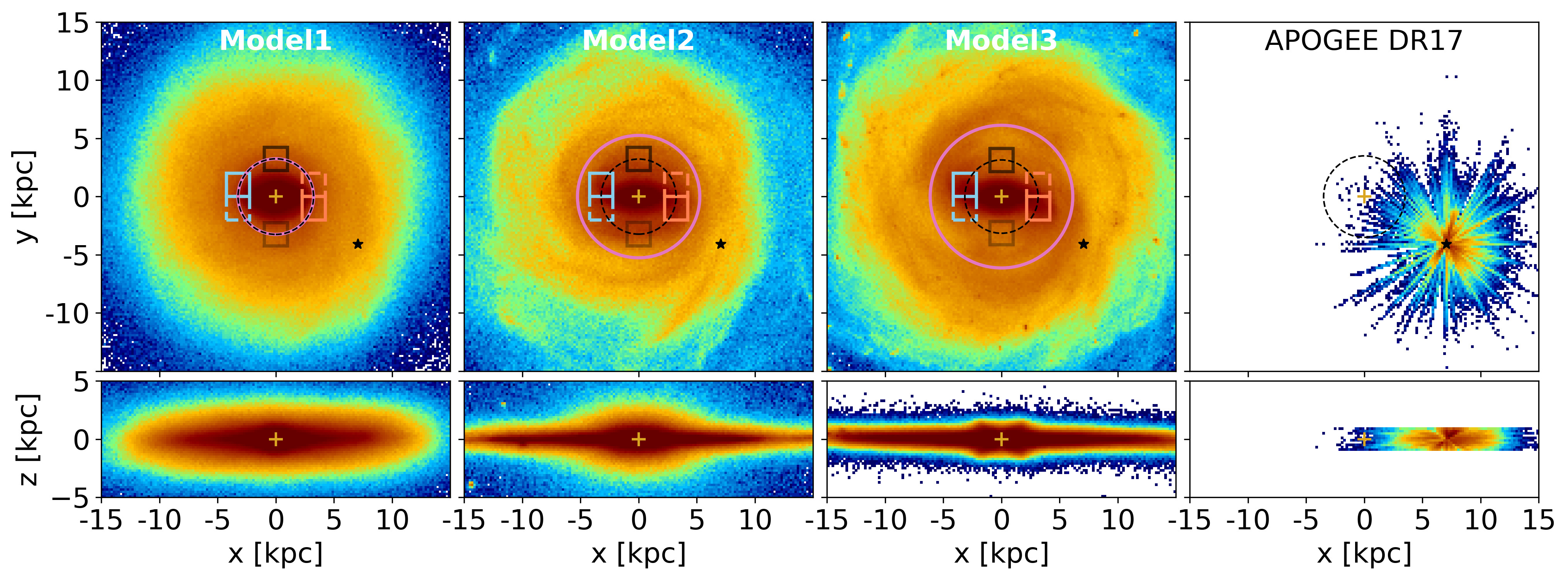}
        \caption{Face-on (top row) and edge-on (bottom row) density map of the three models (from left to right) at the last snapshots, and the APOGEE DR17 sample used (right). The dashed black circle indicates the bar length, while the pink circle indicates the bar's CR. They coincide for Model1. The black star indicates the location of the (simulated) Sun. Rotation is in the counterclockwise direction. The $2\times2$ kpc$^{2}$ squares show the regions which will be studied in Figs. \ref{fig:migr-std_time_limitbar}, \ref{fig:rbar+migreff}, \ref{fig:model1_sfr_diffR} and \ref{fig:model1_sfr+rbar}. Light blue indicates the left side of the bar, while orange indicates the right side of the bar. The solid squares are on the trailing sides of the bar, while the dashed squares are on the leading side of the bar, although leading and trailing sides will not be differentiated in Sec. \ref{sec:starformation} (only one $2\times2$-kpc$^2$ square per edge). The black and gray color indicates squares along the bar minor axis. These two squares will be stacked together and averaged in Sec. \ref{sec:radmigr}.}
        \label{fig:xyz_allsim+obs}
    \end{figure*}

    \section{Simulations}
    \label{sec:sim}
    
    The three hydrodynamical simulations used in this work are the same as those used by \citet{vislosky2024}. They were chosen due to their similarity to the Milky Way, hosting both a bar and a spiral structure, and having similar disk properties. We examine their evolution over the last $\sim1.5$ Gyr, long enough after the bar formation to avoid its effects on disk dynamics. For consistency, the models are called Model1, Model2 and Model3 as in \citet{vislosky2024}, whose aim was to reproduce the Gaia DR3 disk radial velocity field. Model1 and Model2, simulated in the cosmological context, were also used in \citet{hilmi2020} to study fluctuations in bar parameters in relation to the periodic bar-spiral overlap. Model3 is an isolated galaxy, starting from a pre-assembled stellar disk, with gas and evolving with star formation and chemical evolution.
    
    The first three columns of Fig. \ref{fig:xyz_allsim+obs} show face-on and edge-on views of these models. The bars are aligned along the y=0 axis, and are X-shaped bars for Model1 and Model3, in contrast to Model2 which did not go through a buckling phase. The $2\times2$-kpc$^{2}$ squares show the different regions which will be studied in this paper.
    The initial rotation curves of the three models, assumed to be flat, were $V_{0}^{Model1} = 324$ km/s, $V_{0}^{Model2} = 210$ km/s, $V_{0}^{Model3} = 220$ km/s.
    Despite the identical nomenclature with \citet{vislosky2024}, velocities were scaled slightly differently to match the current estimates of the MW rotation curve at the solar radius $V_{0} = 229$ km/s \citep{eilers2019}, instead of $V_0 = 240$ km/s. 
    Similarly, distances were scaled so that the scale lengths of the three galaxies, initially $h_{d}^{Model1} = 5.6$ kpc, $h_{d}^{Model2} = 5.1$ kpc, $h_{d}^{Model3} = 4$ kpc, would match current estimates of the MW scale length $h_{d} = 3.5$ kpc \citep{BHG2016}. The resulting bar lengths for each model are $R_{bar}^{Model1} = 3.25$ kpc, $R_{bar}^{Model2} = 3.23$ kpc and $R_{bar}^{Model3} = 3.15$ kpc. These values were derived using the $L_{cont}$ method described by \citet{hilmi2020}, who define the true bar length only when the bar is not connected to any spiral, as the distance at which the background-subtracted-density drops to 50\% of the maximum central density. The values used here are the time-median of the true bar lengths measured at each time the bar is disconnected from spirals over the last $\sim$ 1.5 Gyr.
    Scaling of velocities and distances affect the mass of the galaxy according to $GM \sim V^{2}R$ (virial theorem).
    
    For Model3, a pre-assembled disk and bulge, disk stars are indicated so their identification is straightforward. For the cosmological simulations however (as well as the observational data), applying cuts is necessary to isolate them. We considered stars to belong to the disk if (1) their birth $|z_{0}|$ and current $|z|$ distance to the mid-plane are both less than 1 kpc, (2) their birth radii $R_{birth}$ are less than 15 kpc, (3) their current galactocentric radii $R$ are less than 25 kpc, and (4) their eccentricity is less than 0.5 (to remove most stars trapped in the bar). Some disk stars have $z>1$ kpc and are therefore removed from the analysis. However, including them did not affect our conclusions.

    The meridional motion of stars can be approximated as the combination of a perfectly circular orbit at so-called "guiding radius" $R_{g}$ (ideal, stable, symmetric case) and an epicyclic motion (radial oscillations) around $R_{g}$ \citep{BT08galdyn}. Thus, in the general case, two types of changes can affect the orbit: change of guiding radius, called radial migration, and/or an increase of epicyclic amplitude, called kinematic heating which can be related to an increase of the eccentricity of the orbit. For both simulations and the data, we estimated the stars' guiding radii $R_g$ and eccentricities $e$ using the following expressions:
    \begin{center}
        \begin{equation}
            R_{g} = R \frac{V_{\phi}}{V_{0}} = \frac{L_{z}}{V_{0}}
        \end{equation}
        \begin{equation}
            e= \sqrt{\frac{V_{r}^{2} + 2\left(V_{\phi} - V_{0}\right)^{2}}{\sqrt{2}V_{0}}},
        \end{equation}
    \end{center}
    where $R$ is the instantaneous galactocentric radius of the stellar particle, $V_{r}$ and $V_{\phi}$ are the radial and azimuthal galactocentric velocities, and $V_{0}$ is the rotation curve. Even though rotation curves are significantly increasing with radii in the central-most parts of disk galaxies, they quickly flatten and we therefore approximate $V_{0}$ to a constant. As a result, guiding radii of the innermost stars could be underestimated. However, dividing $V_0$ by 2 for stars at $R<3$ kpc did not impact our conclusions.
    The eccentricity expression above was taken from \citet{arif2006_ecc} for a flat rotation curve, as we also assume in this work, and is reliable for eccentricities up to 0.5. As mentioned earlier, stellar particles with $e>0.5$ were excluded from our analyses.

    The APOGEE DR17 sample has a complex selection function as can be seen at least in the radial dimension in Fig. \ref{fig:xyz_allsim+obs}. Therefore, to properly compare the last snapshots of the simulations to the observations of APOGEE as will be done in Sec. \ref{subsec:impactSNd}, it is necessary to take these selection biases into account, as well as the measurement uncertainties. Uncertainties were introduced as Gaussians except for birth radii, for which the error strongly depends on age. We split the stars into 1 Gyr monoage bins, and fit the $R_{birth}$ error distributions from APOGEE DR17, which we then implemented for the same age populations in the simulations. For the selection effects, we used rejection sampling for the same age bins, to sample the simulations so that the radial distributions of each simulated age population would be the same as that of the corresponding observed populations. The number of stars for each age bin was kept equal to that in the APOGEE sample, which helps to address selection biases based on stellar evolutionary stage. Details are given in Appendix \ref{selection_effects}.    

    \subsection{Model1}
    \label{subsec:Model1}
    
    The first cosmological model we used is model g2.79e12, a zoom-in version of a Milky Way-mass galaxy taken from the Numerical Investigation of a Hundred Astronomical Objects (NIHAO-UHD) simulation suite \citep{wang2015_nihao}. First presented by \citet{buck2018_sim}, it hosts both a bar and a spiral structure with $\sim$10\% overdensity \citep[ratio of the amplitude of the m=1, 2, 3 and 4 components to the m=0 Fourier component of the stellar density, see][]{vislosky2024}. Hence it has been extensively used due to its similarity to the Milky Way \citep{buck2019a, buck2020, sestito2021, lu2022a, lu2022b, buck2023, wang2023, buck2023}. The simulation is based on the smoothed particle hydrodynamics solver \textsc{GASOLINE2.0} \citep{wadsley2017_gasoline2}. Star formation and feedback are implemented following the prescriptions of \citet{stinson2006_SF+feedback} and \citet{stinson2013_feedback}, who searched for simulation parameters that best reproduced the Kennicut-Schmidt law and the stellar mass - halo mass relation, respectively.

    The simulation parameters can be found in \citet{buck2019a}, the resolution is reported here : $\sim 5.2 \times 10^{6}$ dark matter particles ($5.141 \times 10^{5}$ M$_{\odot}$/particle), $\sim 8.2 \times 10^{6}$ stellar particles ($3.13 \times 10^{4} \text{M}_{\odot}\text{/particle}$, total $\text{M}_{star} = 1.59 \times 10^{11} \text{M}_{\odot}$), and $\sim 2.2 \times 10^{6}$ gas particles ($9.38 \times 10^{4} \text{M}_{\odot}$/particle, total $\text{M}_{gas} = 1.85 \times 10^{11} \text{M}_{\odot}$).

    \subsection{Model2}
    \label{subsec:Model2}
    
    The second galaxy is model g106 of the zoom-in suite of 33 hydrodynamical simulations of \citet{martig2012}. The zoom-in technique \citep{martig2009} consists in extracting the merger and accretion histories of a dark matter halo simulated in a cosmological context \citep[here from][]{teyssier2002}, and then resimulating these histories at high resolution including baryons. The zoomed-in simulation used the Particle-Mesh code from \citet{bournaud2002,bournaud2003}. An extensive description of the simulation technique is given in \citet{martig2012, martig2014a}, but we report some parameters here. Star formation is implemented using the Schmidt-Kennicutt law \citep{kennicutt1998} with an exponent of 1.5 and a 2\% efficiency. They set the threshold for star formation to 1 atom per cubic centimeter (i.e., 0.03 M$_{\odot}$ pc$^{-3}$). Similarly to Model1, this galaxy hosts both a central bar and a $\sim$20-25 \% overdense spiral structure \citep[ratio of the amplitude of the m=1, 2, 3 and 4 components to the m=0 Fourier component of the stellar density, see][]{vislosky2024}. It has also been widely used for its resemblance to the MW \citep{kraljic2012, minchev2013, martig2014a, martig2014b, carrillo2019, hilmi2020}.
    
    The mass resolution for dark matter particles is $3 \times 10^{5} \text{M}_{\odot}$, for stellar particles $7.5 \times 10^{4} \text{M}_{\odot}$ and for gas particles $1.5 \times 10^{4} \text{M}_{\odot}$. The total stellar mass and dark matter mass within the optical radius (25 kpc) are $\sim 4.3 \times 10^{10} \text{M}_{\odot}$ and $\sim 3.4 \times 10^{11} \text{M}_{\odot}$ respectively.

    \subsection{Model3}
    \label{subsec:Model3}
    
    Model3 is an N-body simulation which contains a pre-assembled stellar disk, dark matter halo, and bulge with mass profiles as presented in \citet{khoperskov2022_sim}, to which a gas component was added, including star formation, feedback and enrichment models. A full description was given by \citet{vislosky2024}. It simulates 3 Gyr of evolution of these pre-assembled components.
    Star formation is implemented with an efficiency of 0.05, and occurs in a gaseous cell when: i) gas mass is > $2 \times 10^{5} \text{M}_{\odot}$, ii) the gas temperature is less than 100 K, and iii) if the cell is part of a converging flow.

    Model3 also hosts both a bar and a $\sim$ 20-25 \% overdense spiral structure \citep[ratio of the amplitude of the m=1, 2, 3 and 4 components to the m=0 Fourier component of the stellar density, see][]{vislosky2024}.

    This model contains $5\times10^6$ dark matter particles of mass $1.25\times10^5 \, M_{\odot}$ and $6\times10^6$ stellar particles of mass $7.5\times10^3 \, M_{\odot}$. The total gas mass is $1.5 \times 10^{10} M_{\odot}$.

\section{APOGEE DR17 sample}
    \label{sec:apogee}
    To compare the last snapshots of our simulations with MW observations, we used data provided by the catalog of the Sloan Digital Sky Survey IV \citep[SDSS-IV]{blanton2017} APOGEE's seventeenth data release \citep{majewski2017, abdurro'uf2022}. Stellar parameters and abundances are derived from the data of two high-resolution spectrographs in both the northern and southern hemispheres using the APOGEE Stellar Parameters and Chemical Abundance Pipeline \citep[ASPCAP]{holtzman2015, garciaperez2016, jonsson2020}. We used stellar kinematics from the astroNN catalog \citep{mackereth+bovy2018, leung+bovy2019}, and the uncorrected Sharma model ages from the \texttt{distmass} catalog \citep{imig2023, stone-martinez2024}. We used the birth radii derived by \citet{ratcliffe2024c} for red giant branch stars that belonged to the disk (|z| < 1 kpc, $e$ < 0.5, |[Fe/H]| < 1) with small uncertainties (Fe/H]$_{err}$ < 0.05 dex, unflagged [Fe/H], log$_{10}$Age$_{err;upper}$ > 0).
    Stars with unreliable masses, and therefore ages, because they lie outside the \texttt{distmass} training set were also removed (flagged with BITMASK=2). These selections resulted in a sample of 162,784 disk stars, which is shown in the rightmost panel of Fig. \ref{fig:xyz_allsim+obs}.

\begin{figure*}
        \centering
        \includegraphics[width=17cm]{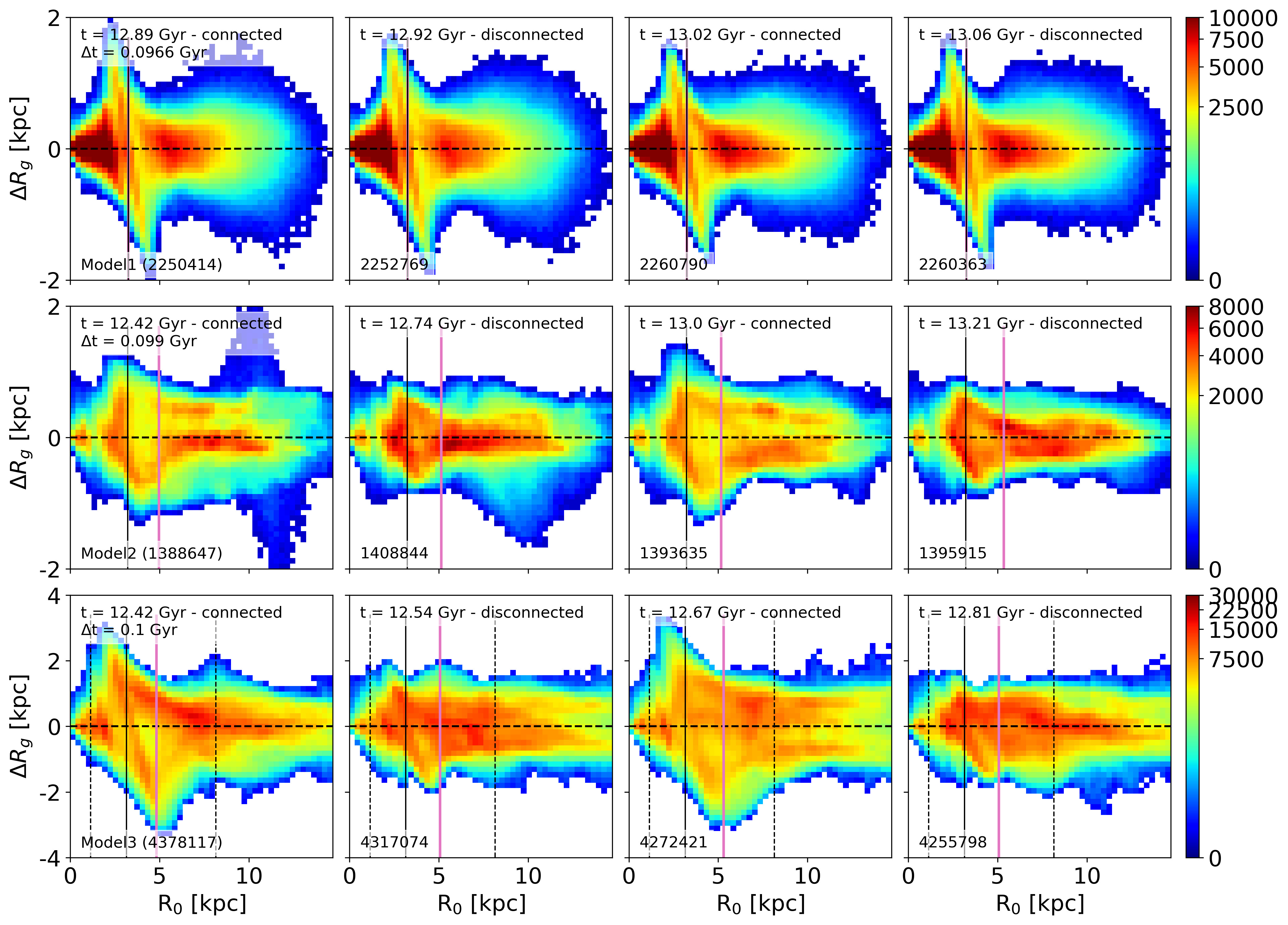}
        \caption{Change of guiding radius over $\Delta t = 0.1$ Gyr vs initial guiding radius for Model1, Model2 and Model3 stars from top to bottom.
        Each column represents a different initial time, increasing from left to right as indicated in the top of each panel. These times were chosen to be at moments when the bar and the spiral are connected or disconnected. The dashed horizontal black line shows $\Delta \text{R}_{\text{g}}=0$, (i.e., no migration). All points above (below) this line represent stars that migrated outward (inward). The solid black line indicates the bar length, while the pink line indicates the bar's corotation radius. They coincide for Model1. The vertical dashed lines in the bottom panel indicate the radial range over which the azimuthal variations of migration are investigated in Fig. \ref{fig:c002_drg_phi}.
        Panels corresponding to bar-spiral connection show a very prominent ridge at the bar end, while this ridge is much less pronounced at bar-spiral disconnection times, except for Model1 where the weak spirals diminish this effect.}
        \label{fig:drg-r0}
    \end{figure*}

    \begin{figure*}
    \centering
        \includegraphics[width=17cm]{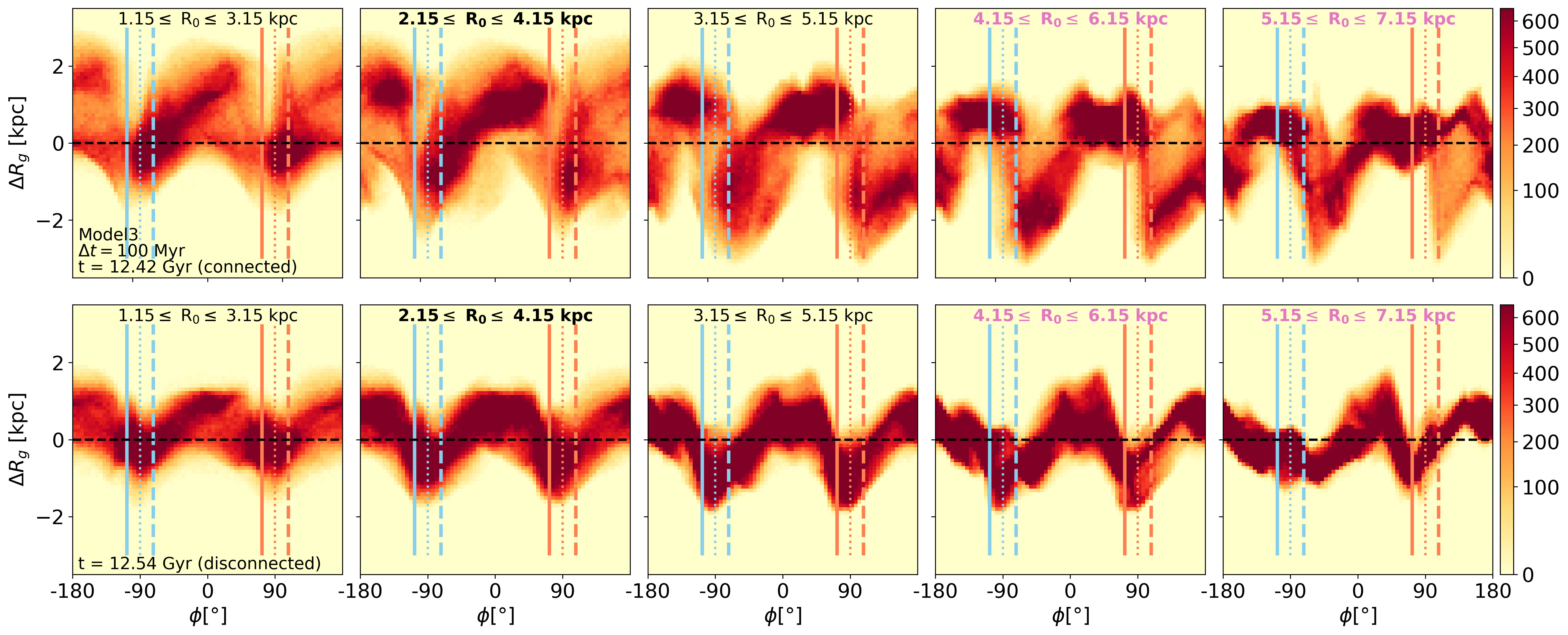}
        \caption{Two-dimensional histograms of the change in guiding radius over $\Delta t = 0.1$ Gyr with respect to the initial azimuthal angle for stars in Model3 initially in different 2-kpc-wide radial bins. Each column is for a different radial bin, going further out from left to right as indicated in each panel. 
        The bar ends ($R_{bar} = 3.15$ kpc) is encompassed in the second column and is indicated in bold, while the corotation ($R_{CR} = 5.3$ kpc) is at the border of the fourth and fifth bins and is indicated in pink.
        The left and right sides of the bar are shown respectively by the light blue and orange vertical dotted lines. The solid lines indicate the trailing side of the bar edges, while the dashed lines indicate their leading sides. Those lines are located at the edge of their associated square regions from Fig. \ref{fig:xyz_allsim+obs}.
        The different rows represent the same quantities at different times of the disk evolution, corresponding to the two first columns of Fig. \ref{fig:drg-r0}, thus showing a time at which the bar is connected to a spiral (first row), as well as a time when the bar is not connected to any spiral (second row). The ($\phi - \Delta R_g$) plane of Fig. \ref{fig:c002_drg_phi} is divided into $5^{\circ} \times 0.1$ kpc bins and the number of stars in those bins is given by the colorbar, with contribution only from stars present throughout the whole period. The full temporal evolution over the last 1.4 Gyr of evolution is available as an \href{https://cloud.aip.de/index.php/s/Ht9A9krNrkGRkLz}{online movie}. 
        %Within and at the bar radius (two innermost bins), migration between the bar edges is outward and stronger than along the bar azimuth where migration is inward. When the bar and a spiral are connected (first row), this tendency is exacerbated: stars migrate even further away than when the bar is not connected to a spiral (second row). From $R_{g,0}\sim 5$ kpc, this behavior along $\phi$ diminishes, indicating a different driver of migration.
        }
        \label{fig:c002_drg_phi}
    \end{figure*}

    \begin{figure*}
        \centering
        \includegraphics[width=17cm]{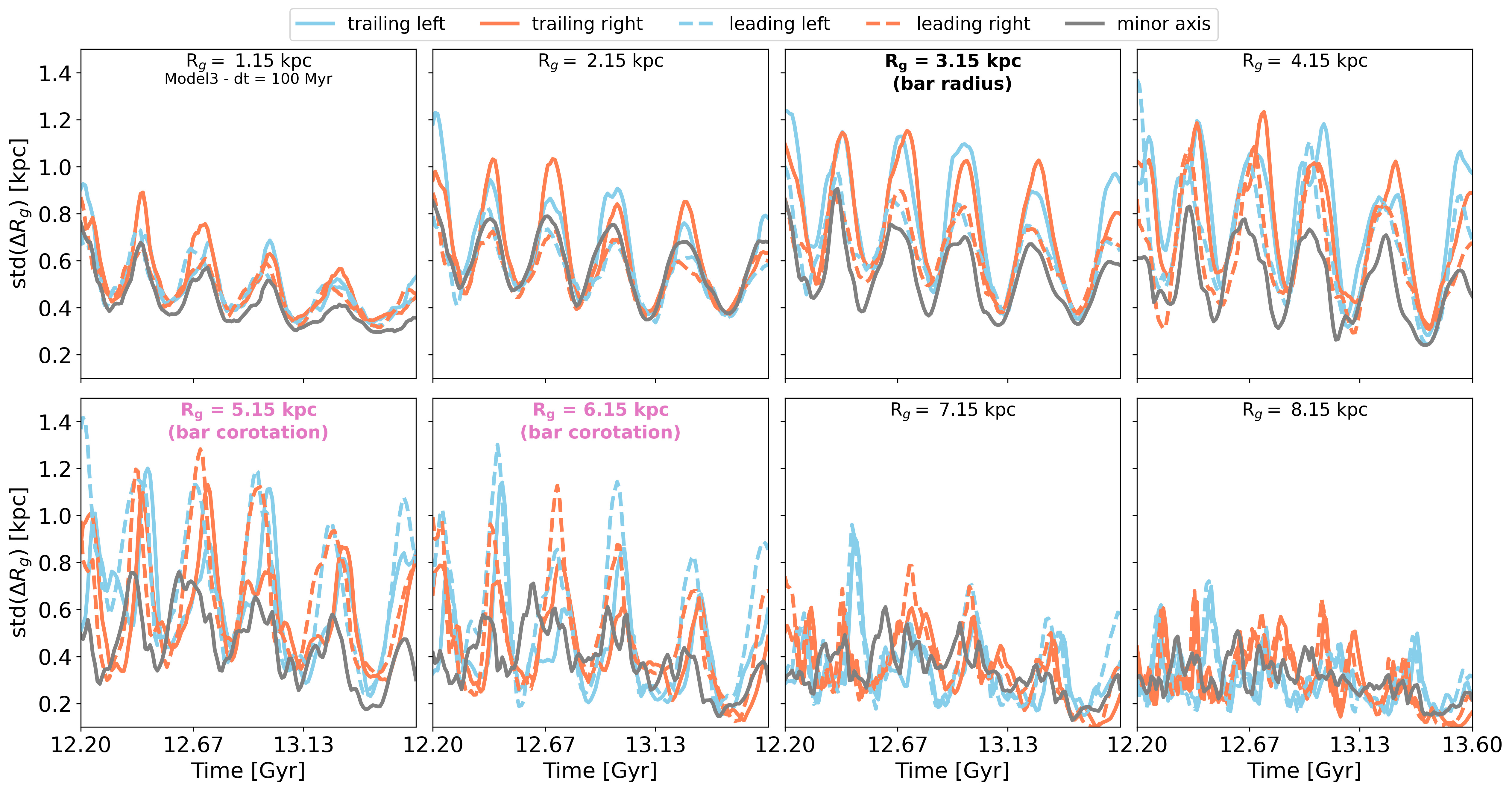}
        \caption{Time evolution of migration strength from different galactic disk radii in Model3. In each panel, the different curves correspond to the migration strength (i.e., how far stars found in this region at some time $t$ migrate after $\Delta t$) of stars originally found in each of the squares shown in Fig. \ref{fig:xyz_allsim+obs}. Light blue and orange indicate respectively the left and right sides of the bar, while solid and dashed represent respectively the trailing and leading sides of the bar. The gray curve is the mean migration strength along the bar minor axis (i.e., the two gray squares of Fig. \ref{fig:xyz_allsim+obs}). From left to right, these squares are shifted horizontally (light blue and orange) and vertically (gray), along the bar major and minor axis respectively, to probe different radii.
        %In the very inner disk (top leftmost panel) the fluctuations' amplitude is relatively small, as spiral arms do not reach this region efficiently and many stars along the bar major axis (blue and orange) are probably either trapped in the bar, or on eccentric orbits. From 2 kpc and up, the fluctuations amplify with the same period out to 5 kpc. After that, the time evolution of migration strength becomes more chaotic, which is likely to be due to the bar losing influence at those distances.
        }
        \label{fig:migr-std_time_limitbar}
    \end{figure*}

    \begin{figure*}
    \centering
        \includegraphics[width=18cm,keepaspectratio]{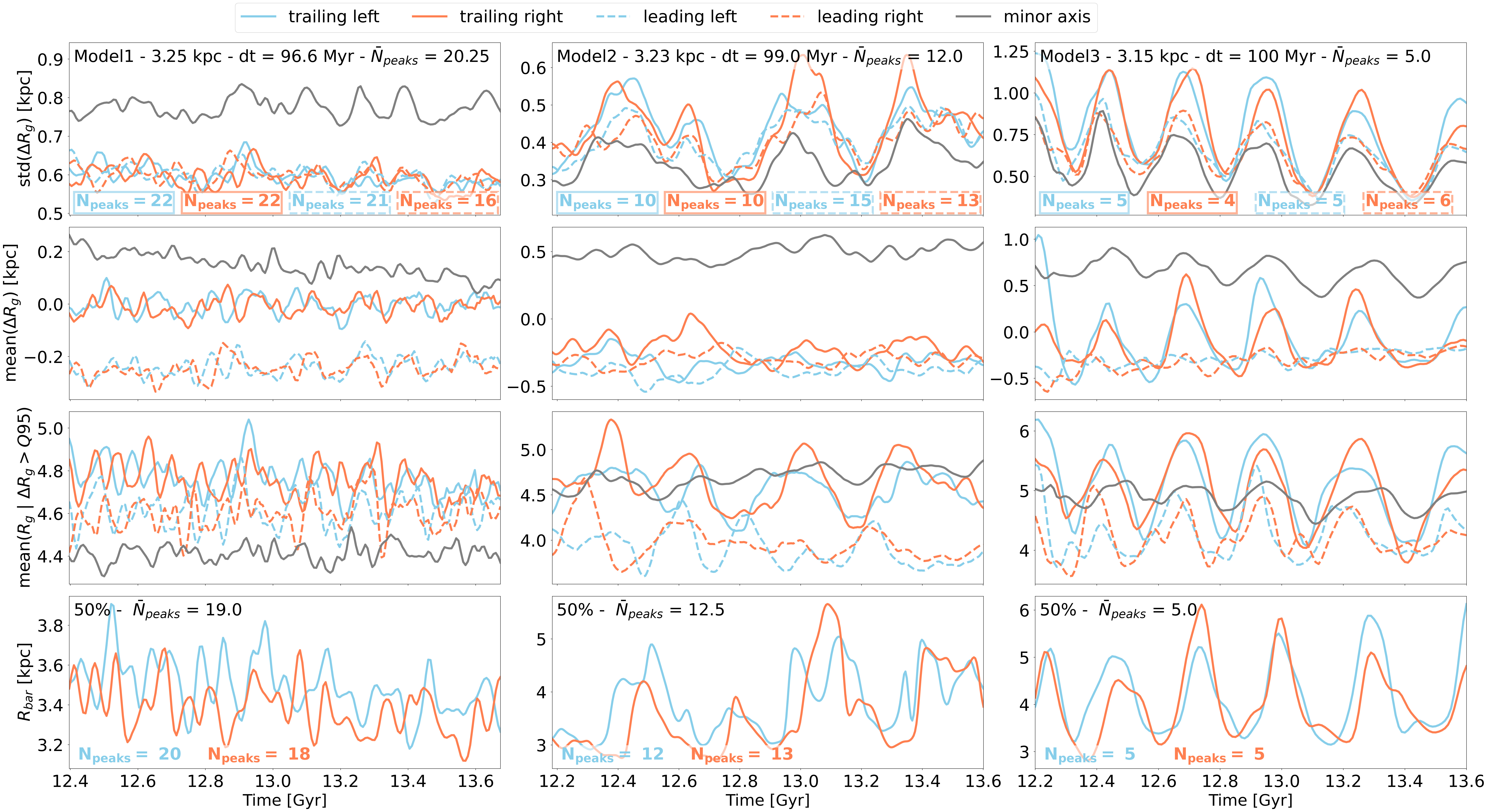}
        \caption{Time evolution of different migration statistics compared to bar length time evolution, for the three models (from left to right). Different azimuths around the bar radius are scanned using the six $2\times2$ kpc$^{2}$ square regions shown in Fig. \ref{fig:xyz_allsim+obs}. The gray curve is the mean statistical quantity over the two squares placed along the bar minor axis. The leading and trailing, left and right sides of the bar major axis are represented by the solid and dashed, light blue and orange curves. The number of peaks of these curves is indicated as $N_{peaks}$ in the top and bottom panel, $\bar{N}_{peaks}$ is the mean $N_{peaks}$ over the different curves.
        \textit{First row:} Standard deviation of the change in guiding radius over $\Delta t = 100$ Myr experienced by all stars located in the six different square-shape regions. %Bar trailing side stars migrate further away than bar leading side stars. Systematic fluctuations are seen in all models.
        \textit{Second row:} Same as top row, but for the mean change in guiding radius, which slows the net angular momentum transfer. % Stars along the bar minor axis migrate outward ($\Delta R_{g} > 0 $). Stars on the leading side of the bar seem to have net inward migration, while stars on the trailing side seem to move either in or out. The same regular fluctuations are found, particularly on the trailing sides of Model1's and Model3's bar. 
        \textit{Third row}: Mean guiding radius after $\Delta t = 100$ Myr for the 5\% stars which migrate the most. %The final $R_{g}$ of stars from the bar ends has the same regular fluctuations seen in the top two rows. The most extreme migrators can reach radii of 6 kpc.
        \textit{Last row} : Time evolution of the bar half-length. The peaks correspond to moments when the bar is connected to a spiral arm. The maxima in the migration strength, or their frequency, coincide with those of the bar half-length, indicating they are a result of the interaction between the bar and spiral arms.}
        \label{fig:rbar+migreff}
    \end{figure*}
    
    \section{Radial migration at the bar radius}
    \label{sec:radmigr}

    \subsection{Time evolution of stellar migration}

    A common way to visualize migration is plotting the change in guiding radius ($\Delta R_{g} = R_{g}(t_{1}) - R_{g}(t_{0} < t_{1})$), versus the initial guiding radius ($ R_{0} = R_{g}(t_{0})$) for a given time period $\Delta t = t_{1}-t_{0}$. From left to right, Fig.~\ref{fig:drg-r0} probes the time evolution of the efficiency and direction of migration over a fixed time interval of $\Delta t = 0.1$ Gyr. The ($R_{0} - \Delta R_{g}$) plane is divided into $0.3 \times 1.7$ kpc$^{2}$ bins for Model1 and Model2 (top and middle panel) and $0.3 \times 2.7$ kpc$^{2}$ bins for Model3 (bottom panel). The number of stars in those bins is given by the colorbar, with contribution only from stars present throughout the whole period. Each panel corresponds to a particular time in the evolution of one of the three models. The times were chosen to be when the bar is either connected or disconnected to a spiral, as indicated in each panel. The black horizontal dashed line is at $\Delta R_{g} = 0$ (i.e., no migration). Every particle above (below) this line migrated outward (inward). The black vertical solid line indicates the bar end, while the pink line indicates the corotation radius (CR). We note that the CR and the bar length overlap in Model1, revealing a fast bar.
    
    A prominent ridge is visible at the bar end in all models, indicating strong ongoing migration at these radii, even though this is inside the bar's CR for Model2 and Model3. Particularly striking in Model3, but also visible in Model2, is how the amplitude of this ridge varies with time. It is most pronounced when the bar connects to a spiral, and almost disappears when they disconnect. The ridge at the bar end remains prominent at all times for Model1. This model has a much weaker spiral structure, therefore the effect of the bar-spiral overlap is less visible here. We note that the ($R_0 - \Delta R_g$) plane, or its equivalent ($L_0 - \Delta L$), has been shown multiple times, but almost always for larger time intervals \citep[e.g.,][]{Roskar2008_migration, minchev2011_res-overlap, halle2018, Khoperskov2020_barres-sweep}. Using a time interval larger than a migration burst due to a specific event can result in blurring the structure (see Fig. \ref{apdx:drg_r0}). This is the first time that such periodic fluctuations in the migration strength, measured as a change in guiding radius, have been shown.

    \paragraph{}Plotting $\Delta R_{g}$ against $R_{0}$ as in Fig. \ref{fig:drg-r0} allows us to study how migration strength varies with radius, but not how it varies with azimuth, i.e angle relative to the bar major axis. Figure \ref{fig:c002_drg_phi} shows two-dimensional histograms of the change in guiding radius, $\Delta R_g$, as a function of azimuthal angle, $\phi$, which has resulted in a time period $\Delta t = 0.1$ Gyr for Model3 stars in different radial bins. For simplicity, most results in the rest of the subsection will focus on Model3. Differences with Model1 and Model2 will be discussed below.
    Figure \ref{fig:c002_drg_phi} is the azimuthal expansion of the region between the two vertical black dashed lines in the bottom row of Fig. \ref{fig:drg-r0}.
    Different columns show different guiding radius bins of width 2 kpc, spanning a region from 1.15 kpc (leftmost) to 7.15 kpc (rightmost). The bins overlap with one another by half the bin width. The bar ends are encompassed in the radial bin indicated in bold (second column), while its CR is at the border of the two radial bins indicated in pink (the two last columns). The two rows correspond to the first two columns of the last row of Fig. \ref{fig:drg-r0}, that is times when the bar is either connected (first row) or disconnected (second row) to a spiral.
    We can see that out to $R_g\sim 6$ kpc a large number of stars initially located at the bar ends appear to mostly migrate slightly inward ($\Delta R_{g}<0$). Although we removed high-eccentricity orbits ($e>0.5$), this is likely due to them being trapped into the bar-supporting $x_1$ orbits \citep[e.g.,][]{contopoulos80a}, making it hard for them to escape the bar. Between the bar ends and along the bar trailing side however, stars migrate mostly outward to a variety of radii. Particularly, we note that stars initially at the bar radius (second column) can gain almost 3 kpc in radius in just 0.1 Gyr, reaching radii around 6 kpc, where they could once again be pushed away to, for example, the SNd by other migration inducing agents.
    
    Looking more closely at the bar edges in Fig. \ref{fig:c002_drg_phi}, asymmetry can be noticed on the leading and trailing sides. $\Delta R_g$ seems to increase faster on the trailing side (solid lines, decreasing $\phi$) than on the leading side (dashed lines, increasing $\phi$), indicating larger migration on the former. This tendency was observed in other numerical works \citep[e.g.,][]{ceverino2007, petersen2019}, and its effects are also potentially seen in observed face-on barred galaxies \citep{neumann2024}. It can be explained by the positive and negative torques applied respectively on the trailing and leading sides of the bar.
    
    Furthermore, when the bar and the spiral are disconnected (bottom row of Fig \ref{fig:c002_drg_phi}), the variations in migration along $\phi$ are the same for all radii, with minima along the bar axes. However, when looking at times of connection of the bar with a spiral (top row of Fig \ref{fig:c002_drg_phi}), both the amplitude and phase of $\Delta R_g(\phi)$ seem to change with radius.
    Going from innermost to outermost radii (left to right in the top row of Fig. \ref{fig:c002_drg_phi}), $\Delta R_g$ drops with radius, being mostly positive inside the bar radius (first column), and mostly negative outside the bar radius (from third panel). This trend is at the origin of the ridge at the bar radius seen in Fig. \ref{fig:drg-r0}, where net migration is indeed outward inside the bar radius, and inward outside the bar radius. Moreover, the minima in $\Delta R_g$ are exactly at the bar ends for all radii when the bar is not connected to any spiral (bottom row). This is true only inside the bar radius, a region which spirals do not reach, when the bar is connected to a spiral (top left panel). At this moment of bar-spiral overlap, the minima in $\Delta R_g$ shift in azimuth as radius increases (from left to right in top column of Fig. \ref{fig:c002_drg_phi}), going from the bar azimuths and moving to the leading side of the bar.
    Therefore, Fig. \ref{fig:c002_drg_phi} suggests that not only the torques vary with time as a result of bar-spiral overlap (causing stronger migration when connected), but torques seem to also shift in space, moving the zones of inward/minimal migration from exactly at bar end when bar and spiral are disconnected (bottom row), to along the leading side of the bar when bar and spiral are connected (top row).
    This behavior is verified across the 1.4 Gyr of evolution studied here, as can be seen in the \href{https://cloud.aip.de/index.php/s/Ht9A9krNrkGRkLz}{online movie}.

    \paragraph{}To visualize how migration strength along azimuth varies with time near the bar ends, we looked at the standard deviation of the distribution of $\Delta R_{g}$ along the initial azimuthal angle $\phi_{0}$ in different radial bins. Because mean migration is around 0 kpc, the standard deviation of $\Delta R_{g}$ indicates how far stars typically migrate, by ignoring the canceling effect of inward ($\Delta R_{g} < 0$) and outward ($\Delta R_{g} > 0$) migrators. We therefore define the migration strength as the standard deviation of $\Delta R_{g}$. Figure \ref{fig:migr-std_time_limitbar} shows the time evolution over 1.4 Gyr of migration strength, still over $\Delta t = 0.1$ Gyr, for the radial bins of the different square regions shown in Fig. \ref{fig:xyz_allsim+obs}, but centered at increasing radii from top left to bottom right, following the radial bins of Fig. \ref{fig:c002_drg_phi}. Each curve in Fig. \ref{fig:migr-std_time_limitbar} is the migration strength associated to the same-color and same-linestyle squares. The gray curve is the mean migration strength along the two bar minor axis squares stacked together. The squares centered at the bar radius are those shown in Fig \ref{fig:xyz_allsim+obs} and the corresponding migration strength evolution is the third panel of the first row, where the radial bin is indicated in bold. The other panels correspond to a horizontal (vertical) translation of these squares for the light blue and orange (gray) curves.
    First comparing the leading and trailing sides of the bar, we notice that out to the bar radius (up to third and fourth panels), regions along the trailing side of the bar show more migration than along the leading side of the bar, as noticed in Fig. \ref{fig:c002_drg_phi}. This is no longer true from the CR, around 5 kpc (bottom left panel). The gray curve is almost always lower than the colored curves, indicating less variety in migration along the bar minor axis (all minor axis migrators reach more similar radii). Showing the two opposite sides of the bar minor axis separately did not change this trend. Despite these differences, all curves show periodic fluctuations on similar frequencies in the migration strength around the bar up to just outside the bar radius (top right panel), generalizing observations from Fig. \ref{fig:drg-r0} and \ref{fig:c002_drg_phi} to a longer timescale. Around CR, significant fluctuations are still present. This could be explained by the fact that the bar-spiral overlap also makes the bar pattern speed oscillate, which in turn affects its resonances as well \citep{hilmi2020}. Outside the CR, fluctuations are still present, but are much less regular as another migration agent must have taken over.
    The slight decrease at all radii of the peaks in migration strength with time is likely because the disk gets hotter and thus less sensitive to perturbations like the bar and the spiral. We note that these oscillations in guiding radius could also be the result of libration around the fourth and fifth Lagrange points, as was shown to occur around the bar's corotation \citep{ceverino2007, BT08galdyn, haywood2024, khoperskov2024c}. This effect, however, does not produce a continuous increase in $\Delta L$ (or $\Delta R_g$) with time \citep[see, e.g.,][, Fig. 2]{minchev2010}. In contrast, we find that the mean $\Delta R_g$ is overall positive (see Fig. 5), suggesting that migration is indeed happening, on top of libration.

    \paragraph{} \citet{hilmi2020} showed that the bar appears longer when it overlaps with a spiral mode. To verify that the boosts in migration seen in Fig. \ref{fig:migr-std_time_limitbar} are related to bar-spiral overlap, we compared them with the time evolution of the bar length. Focusing back on the bar region, in Fig. \ref{fig:rbar+migreff} we present several statistics of the population of stars found at a time $t$ in one of the 5 regions around the bar shown in Fig. \ref{fig:xyz_allsim+obs} : the standard deviation of their $\Delta R_{g}$ in $\Delta t$, which we call the migration strength ; their mean $\Delta R_{g}$  in $\Delta t$; and the mean $R_{g}$ that their 5\% most extreme migrators reach after $\Delta t$. Figure \ref{fig:rbar+migreff} compares the time evolution of these quantities to the bar length evolution for our three models. Each column corresponds to one model as indicated in the top panels, while each row shows a different quantity. As for Fig. \ref{fig:migr-std_time_limitbar}, each curve represents the migration statistics of the corresponding region. The top rightmost panel is the same as the top second-to-last panel of Fig. \ref{fig:migr-std_time_limitbar}. Curves in the three first rows are smoothed with Gaussian kernels of standard deviation 5 Myr, 3 Myr and 7 Myr from left to right respectively. Bar lengths curves are smoothed with Gaussian kernels of standard deviation 8 Myr, 9 Myr, and 17 Myr. Smoothing widths are functions of the time resolution of each simulation. Since the bar length measurements are more susceptible to noise than the migration statistics (which are somewhat already averaged), a stronger smoothing is applied to bar length measurements.
    
    All the models show oscillations, but with different frequencies. Model1 has a faster bar, and exhibits more rapid oscillations in its migration statistics for all azimuth around the bar, while the slower bars of Model2 and Model3 correlate with slower fluctuations. This highlights the importance of the bar in producing these migration bursts. 
    We also note the stronger migration in Model3, which exhibits higher values for all migration statistics than in Model1 and Model2. These galaxies, simulated in the cosmological context, could be hotter because of their interaction with other objects, making them less sensitive to bar and spiral perturbations. There could also be additional spurious heating due to the lower resolution in these simulations \citep{sellwood2013, ludlow2021, ludlow2023, wilkinson2023}.

    The std($\Delta R_g$) (first row in Fig. \ref{fig:rbar+migreff}) is higher for trailing sides (solid) than leading sides (dashed) in Model2 and Model3 particularly, suggesting more migration there as seen in Fig. \ref{fig:c002_drg_phi} and Fig. \ref{fig:migr-std_time_limitbar}. This is less clear for Model 1, although we do find some significant difference between leading and trailing sides in the mean change in guiding radius.
    The mean change in guiding radius (second row) allows us to study the direction of migration. The gray curve, representing the bar minor axis, is above zero at all times, indicating a net outward migration. This confirms that these oscillations result from the permanent redistribution of angular momentum, and not pure libration. On the contrary, the leading sides of the bar (dashed) produce negative mean $\Delta R_g$ in all models, again hinting towards negative torques applied on this side of the bar. The trailing sides of the bar (solid light blue and orange) seem to oscillate around zero (except for Model2 where mean$(\Delta R_g)<0$ kpc), suggesting both positive and negative torques successively applied on the stars. The fact that stars on the trailing sides of the bar migrate further away than those on the leading side is confirmed by the mean guiding radius reached by the 5\% most extreme migrators in $\Delta t = 0.1$ Gyr (third row) in all models. This final radius is much higher at all times for the trailing sides, reaching up to 6 kpc. On longer timescales, the stars could even reach larger radii, thus entering the SNd.
    
    The last row of Fig. \ref{fig:rbar+migreff} shows the bar half-length time variations of all models. The bar length is calculated using the $L_{cont}$ method described by \citet{hilmi2020}, defining the bar length as the minimum distance at which the background-subtracted density profile drops by 50\% compared to the radial mean density along the major axis of the bar. This also corresponds to the bar being separated from the spiral. The bar length fluctuates as shown by \citet{hilmi2020, vislosky2024}.
    The peaks in $R_{bar}$ correspond to times when the bar and the spiral are connected. Their frequency coincide quite well with the boosts in the migration strength of all models (first row), and in the mean change in guiding radius along the bar minor axis and the trailing side of the bar, despite these quantities being measured independently and with very different methods. \citet{hilmi2020} performed a Fourier decomposition of this model, and found a reconnection time of $T_{rec} \sim 60$ Myr between the bar and the m = 2, 3 and 4 spiral modes. Over a timescale of 1.3 Gyr as used here, this should result in around 21 peaks, which is very similar to the number of peaks found here, as indicated in the top and bottom panels of Fig. \ref{fig:rbar+migreff}. Peaks are counted using the \texttt{find\_peaks} function from the python library \texttt{scipy}, which identifies peaks as local maxima by comparison of neighboring values. Similarly for Model2, \citet{hilmi2020} found a reconnection time of $T_{rec} \sim 105$ Myr for the bar and the m = 2, 3 and 4 spiral modes. Over the 1.4 Gyr represented here, this would mean around 13 peaks in migration and bar length. Our results are indeed very close to this result, corroborating the link between bar-spiral coupling, and regular migration boosts. In Model3, the bar length and migration peaks are much smoother, and their time offset quite small, so their correlation is more obvious.
    Since all bar parameters, length, strength, and pattern speed, oscillate as a result of this regular coupling, the migration fluctuations can be linked to both the periodic fluctuations in density at the bar ends, when the bar and a spiral arm physically overlap, and from the bar-spiral resonance overlap creating a stochasticity in this region of phase space \citep{minchev2010}. These two processes most likely work in unison.

    \begin{figure*}
        \centering
        \includegraphics[width=18cm]{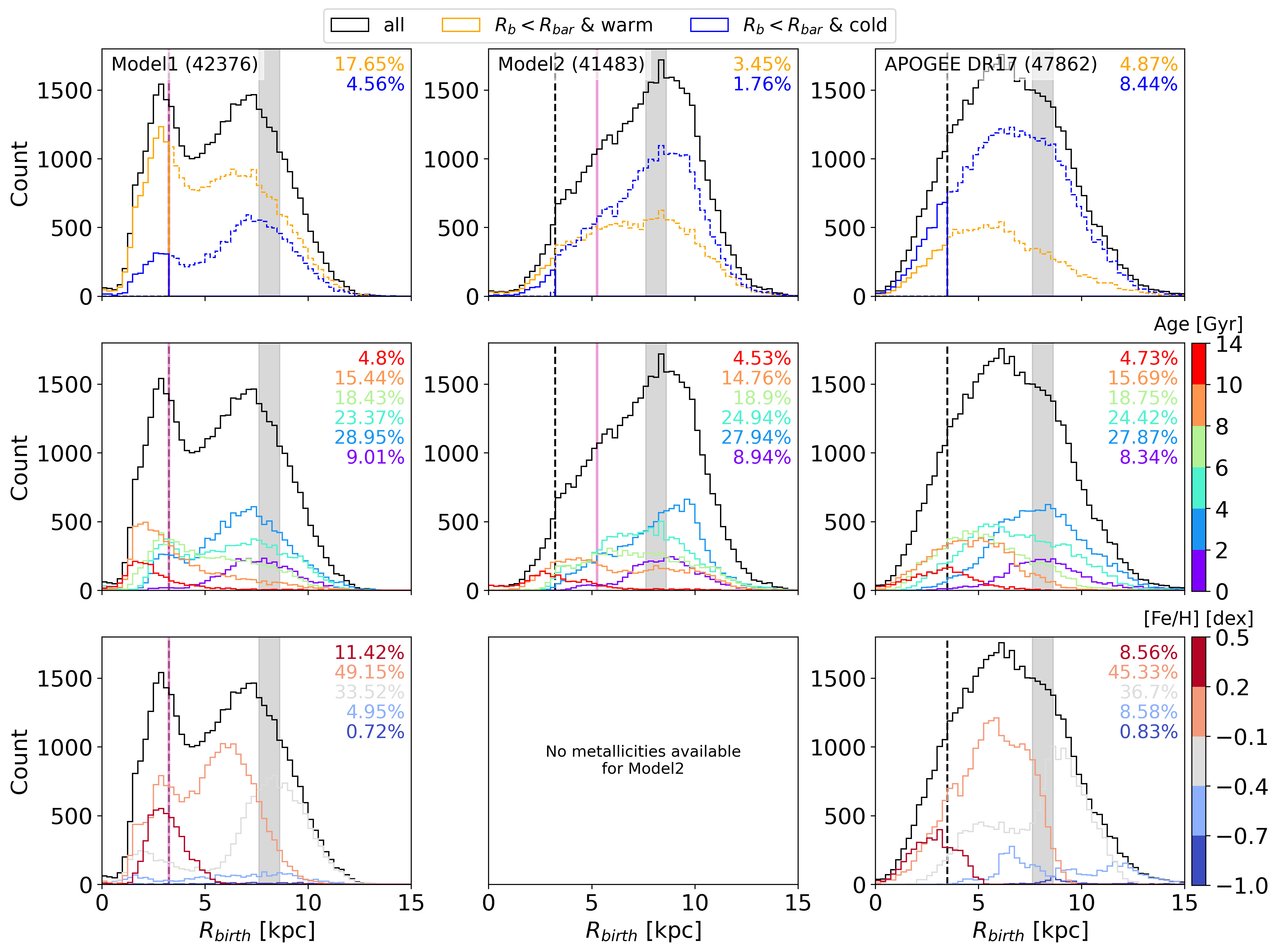}
        \caption{Birth radius distribution of stars found at the solar radius today (inside the gray vertical strip) of, from left to right, the two cosmological models Model1 and Model2, and the APOGEE DR17 red giant sample. The black curves are for the whole samples. In the top row, the blue and yellow curves highlight stars found today on orbits with respectively $e<0.2$ (cold) and $0.2 \leq e < 0.5 $ (warm). The percentages that warm and cold stars born inside $R_{bar}$ represent in the SNd is indicated in the top right of the panels for each galaxy. In the middle row, $R_{birth}$ distributions are shown for different monoage populations. From purple to red, stars get older and older, as indicated in the colorbar. In the bottom row, $R_{birth}$ distributions are shown for different metallicity bins, blue and red representing respectively metal-poor and metal-rich stars, as indicated in the colorbar. %In all panels, the vertical pink and dashed black lines indicate respectively the CR and bar radius of each model. Overall, the youngest stars were born mostly at the solar radius as they did not have much time to migrate. As stars get older and more metal-rich, their birth radius decreases, consistently with an inside-out formation scenario. In all cases, a significant portion of stars from inside 4 kpc reached the solar radius on cold orbits.
        }
        \label{fig:Rb-hist}
    \end{figure*}

    \begin{figure*}
        \centering
        \includegraphics[width=18
cm]{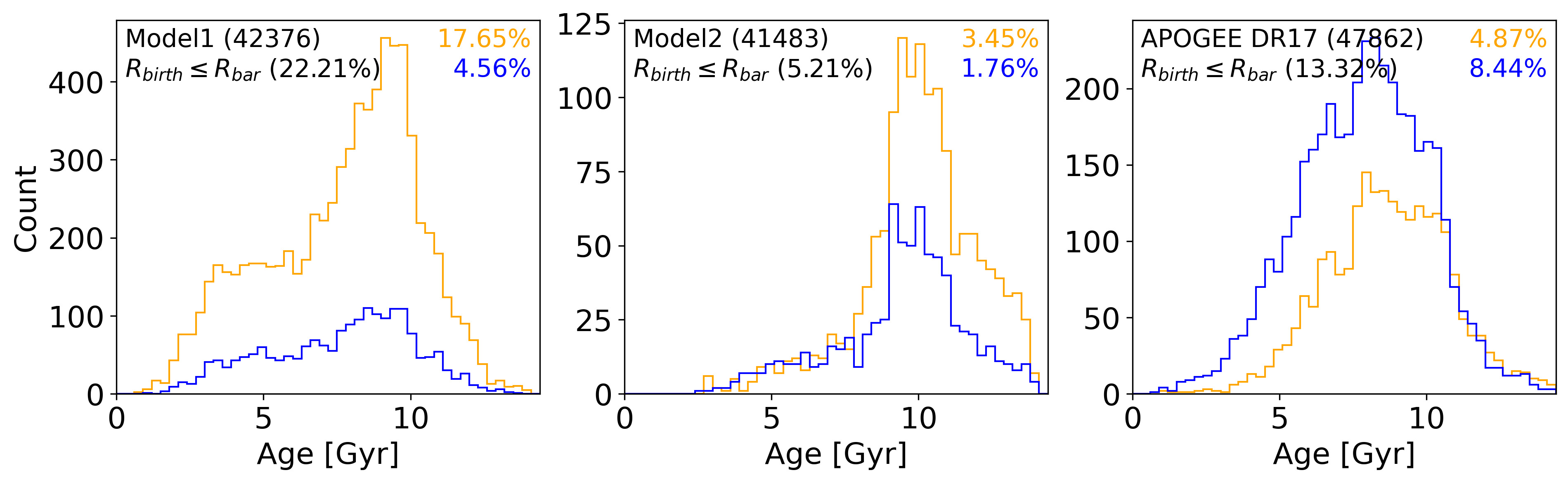}
        \caption{Age distribution of stars found at solar radius today but born inside $R_{bar}$. The two curves follow the same linestyle and color code as in the top row of Fig. \ref{fig:Rb-hist}, solid for stars born inside $R_{bar}$, blue and yellow for kinematically cold and warm stars, respectively. The percentages in the legend are the proportion each eccentricity population with $R_{birth}<R_{bar}$ represents in the whole solar radius sample indicated in gray in Fig. \ref{fig:Rb-hist}. In the three cases, both the cold and warm distributions are mostly old (10 Gyr in the models, 8-9 Gyr in the data) with a tail to younger ages.}
        \label{fig:Age_Rb4}
    \end{figure*}
    
    \subsection{Impact on the solar neighborhood}
    \label{subsec:impactSNd}

    We showed in Fig. \ref{fig:c002_drg_phi} and Fig. \ref{fig:rbar+migreff} that stars in the vicinity of the bar could migrate out to almost twice their radius in just 100 Myr, thus potentially later ending up in the SNd and polluting the local chemistry. To better quantify this effect, we studied the birth radius distribution of stars found today in the SNd of our two cosmological models and the APOGEE data. Model3 is a pre-assembled disc, and can therefore not reproduce the age distribution found in the data. Moreover, it is isolated, and so less realistic than cosmological models. Hence we did not consider it in this subsection.
    
    Figure \ref{fig:Rb-hist} shows the distribution of birth radii of stars found today in the SNd for, from left to right, Model1, Model2, both biased with measurement errors and according to APOGEE monoage radial distributions, as explained in Appendix \ref{selection_effects}, and the APOGEE DR17 red giants sample. The SNd stars are defined as stars with current radius at most 0.5 kpc away from the Sun's radius ($R_{\odot}=8.125$ kpc). This region is indicated by the gray band in the three rows of Fig. \ref{fig:Rb-hist}. In the top row, we show two curves for stars with two different orbital eccentricities today. The blue and yellow are respectively for $e<0.2$, and $0.2 < e < 0.5$ . For simplicity, we call them 'cold', and 'warm' orbits ('hot' orbits would have $e \geq 0.5$). The black dashed vertical line indicates the bar radius of each model. For the MW, we used $R_{bar} = 3.5$ kpc, in accordance with recent measurements \citep{lucey2023, zhang2024a}. Distributions vary from model to model. Model1 shows a bimodal $R_{birth}$ distribution, with a large fraction of stars, both cold and warm, coming from about 1 kpc inside the solar radius, and another significant population born just inside the bar radius. The distributions are smoother in Model2 and in the data, with most stars coming from just inside or at the solar radius, with a slow decrease towards inner birth radii, and a faster one towards the outer disk. This indicates more outward migration than inward migration. The proportions of SNd stars today on cold and warm orbits, and born inside the bar radius go from 1.75 to 8.5\% for cold stars, and from 3.5 to almost 18\% for warm stars. Model2 is the model with the least amount of migrators originating from the bar. These differences could arise from different reasons. First, Model1 has a shorter reconnection time than Model2 \citep[see][]{hilmi2020}, which means that the bar-spiral overlap migration mechanism has had more occasions to send stars away, therefore contaminating the SNd even more. We can also speculate that the bar-spiral reconnection frequency matches better the libration frequency, which could produce more efficient migration. Moreover, the spiral structure in Model1 is weak compared to Model2, which can explain why more stars come from the bar rather than from the disk. Furthermore, the majority of stars born at the bar radius in Model1 are on warm orbits. Since the simulated SNd samples were selected based on stellar instantaneous radii, many of these warm stars could be on their apocenters. Finally, Model1 experienced a massive merger about 4 Gyr ago \citep{buck2023}, which must have heated stellar orbits as well as caused large scale migration. In contrast, the MW's last 8 Gyr have been more quiet \citep{martig2014a, martig2014b, haywood2018_gse, helmi2018_gse, dimatteo2019b, kruijssen2020}. Despite these differences, a significant portion of stars from all models, comes from inside the bar, and is found today on cold and warm orbits.
    
    \paragraph{}To further characterize the stars found in the SNd, we show in the middle panel of Fig. \ref{fig:Rb-hist} the same histograms, but with curves colored by age. The black curve represents the $R_{birth}$ distribution for all stars found in the SNd today. In the middle row of Fig. \ref{fig:Rb-hist}, each colored curve is for an age bin of width $\Delta \text{age} = 2$ Gyr, as indicated in the colorbar. Since the oldest stars are rare, the oldest bin contains all stars older than 10 Gyr to allow for a similar number of stars as in the other age bins (see number of stars in the legend of each panel). The youngest stars (purple) were born mainly in the SNd, because their young age has not yet allowed them to migrate significantly. In Model2 (middle column), some stars born 2-4 Gyr ago seem to have migrated inward, possibly from an encounter with a satellite or outer spiral resonances. Despite this peculiarity, as age increases, the peak in the $R_{birth}$ distribution overall shifts to inner radii as predicted by an inside-out formation scenario. It is therefore mostly the old stars that come from inside the bar.

    In the bottom row of Fig. \ref{fig:Rb-hist}, we show the birth radius distributions of different mono-metallicity populations. From blue to red, the metallicity increases as indicated in the colorbar. We do not have metallicity for Model2, hence the empty panel. In both the data and Model1, as birth radius decreases, stars get more and more metal-rich. In particular, stars born inside the bar radius (vertical black dashed line), are mainly metal-rich. This is in good agreement with several other works attributing the origin of the super metal-rich stars to the inner Galaxy only using chemistry and orbital properties from a variety of Galactic surveys, with some relating it to bar evolution \citep[e.g.,][]{kordo2015_obs-radmigr, feuillet2018, Khoperskov2020_barres-sweep, haywood2024, nepal2024}.
    
    Figure \ref{fig:Age_Rb4} shows in particular the age distribution of the stars born inside the bar and found today in the SNd on cold (blue) and warm (yellow) orbits. This is the age distribution of the stars represented by the solid part of the blue and yellow curves in the top row of Fig. \ref{fig:Rb-hist}.
    As suggested in the middle row of Fig. \ref{fig:Rb-hist}, most of them are old stars, with ages peaking at around 10 Gyr in the simulations, and more around 8 Gyr in the MW. This is not a surprising result since they had the most time to migrate and travel larger distances. In future work, we plan to investigate the time at which those stars migrated to disentangle the possible causes for their migration: internal mechanism, or a major merger event, like Gaia-Sausage Enceladus for the MW \citep{belokurov2018_gse-age, helmi2018_gse, haywood2018_gse}, which should have pushed away a large number of stars. The older stars that did not migrate because of an interaction with an external perturbation must have migrated due to internal processes, like a bar-spiral overlap as studied in this paper, due to a slow down of the bar \citep{Khoperskov2020_barres-sweep}, or an overlap of bar CR and spiral inner Lindbald resonance \citep{minchev2011_res-overlap}. In all galaxies, we also see a long tail of younger stars, which must also have migrated from these internal processes. Figure \ref{fig:Age_Rb4} also shows how the distribution of cold and warm extreme migrators in each galaxy is quite different, with a majority of cold stars in the MW, but a larger proportion of warm stars in the two simulations. As mentioned above, Model1 had a late merger around 4 Gyr ago \citep{buck2023}, much later than the last merger of the MW, which must have heated stars. Moreover, simulated galaxies often have hotter stars compared to observations \citep{sellwood2013, ludlow2021, ludlow2023, wilkinson2023}, which could explain this difference. Despite that, we note that all these stars are on rather cold orbits today ($e<0.5$). Stars with $e>0.5$ were removed, but represented a negligible fraction in the simulated SNds (8\% of total SNd stars in Model1, and less than 1.5\% in Model2). Since orbit cooling is unlikely \citep[but marginally possible under the effects of the bar slowing down, see][] {Khoperskov2020_barres-sweep}, most stars must have migrated without significant heating, as those would be preferentially selected \citep[this is known as "provenance bias", see][]{vera-ciro14}.     The two known migration mechanisms that are capable of migrating stars from the very inner regions to the SNd while maintaining their cold orbits are the bar slow down of \citet{Khoperskov2020_barres-sweep} and the bar-spiral interaction mechanism presented in this paper. Since the bar slow-down is significant only during the formation of the bar, the migration it produces should concern stars approximately as old as the bar, about 8 Gyr according to most recent estimates \citep[e.g.,][]{sanders2024, haywood2024}. Bar-spiral interaction should be the most impactful once the bar is well formed, hence should concern younger stars. Therefore, we argue that, apart from merger-related migration, the oldest most extreme migrators migrated as a result of the bar slowing down during its formation, maintaining (or increasing) their circularity, while the younger extreme migrators reached the SNd on cold orbits in consequence to the bar-spiral interaction. Overlap of resonances could explain the kinematically warmer migrators.

    \paragraph{} To summarize this section, we found periodic oscillations in the migration strength near the bar radius of three different models, one isolated and two in a cosmological context. These oscillations happen on timescales of the bar-spiral beat frequency. We interpret this correlation as a new mechanism for migration, produced by bar-spiral interaction, similarly to the mechanism of transient spirals since the bar radius experiences a time-varying potential due to its interaction with various spiral modes. Oscillations are seen from the bar ILR to outside its corotation. This region is a zone where stars librate around the bar's L$_4$ or L$_5$ Lagrange points. We speculate that migration due to bar-spiral interaction should be the most efficient if the bar-spiral beat frequency matches well the libration frequency of stars. %This is in a similar logic as that of migration near the corotation of transient spirals.
    Migrators from bar-spiral interactions are more efficient when they start from along the minor axis of the bar than along the bar major axis, where a lot of stars are trapped in the bar. In addition, we see stronger outward migration for stars located at the trailing edge of the bar compared to those located at the bar leading edge, which we interpret as torques of opposite sign applied along these two sides of the bar, in accordance with previous studies. Mean migration is positive for stars initially along the minor axis of the bar, confirming that libration is not sufficient to explain those guiding radius oscillations. The 5\% most extreme outward migrators reach radii of 4 to 6 kpc in 100 Myr, suggesting that stars born at the bar radius can reach the SNd when given a bit more time.\\
    To investigate the impact of extreme migrators in the SNd, we compared the simulated SNd at the last snapshots of our cosmological models to APOGEE DR17 data. Looking at the birth radius distribution of these SNd stars, we find that 5 to 23\% of them were born inside the bar radius, and are mostly old and metal-rich. From their age and eccentricity, we deduce that extreme migrators reached the SNd schematically as a result of four different mechanisms: after mergers for old (as old as and older than the merger) warm stars ; from the overlap of bar and spiral resonances for cold and warm stars of any age younger than the bar ; from the propagation of the bar resonances due to its slow down for currently kinematically cold stars with age close to the age of the bar ; from the bar-spiral interaction mechanism presented in this paper for cold stars younger the bar.

    \begin{figure*}
        \centering
        \includegraphics[width=16cm]{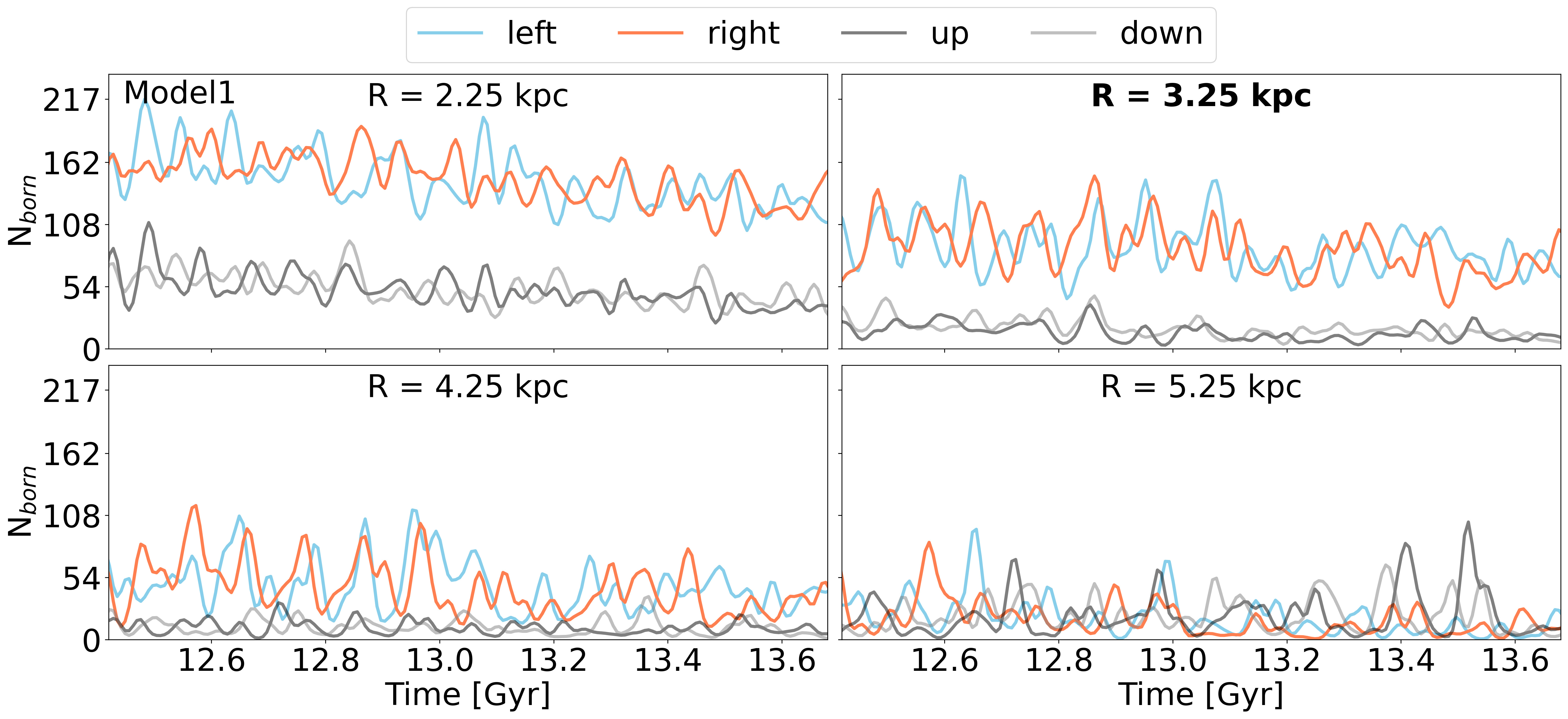}
        \caption{Model1 star formation temporal evolution in $2\times2$ kpc$^{2}$ square regions centered at increasing radii from left to right and top to bottom, spanning radii from 2.25 kpc to 5.25 kpc. The squares of the leading and trailing sides of the bar, shown in Fig. \ref{fig:xyz_allsim+obs}, are merged into one square per edge of the bar centerd on $(x,y)=(\pm R,0)$, light blue and orange indicating respectively left and right side. The upper and lower edges of the bar minor axis are differentiated, using respectively black and gray color. The regions at the bar ends show i) lots of fluctuations, ii) more star formation than along the bar minor axis until just outside the bar radius.}
        \label{fig:model1_sfr_diffR}
    \end{figure*}
    
    \begin{figure*}
        \centering
        \includegraphics[width=18cm]{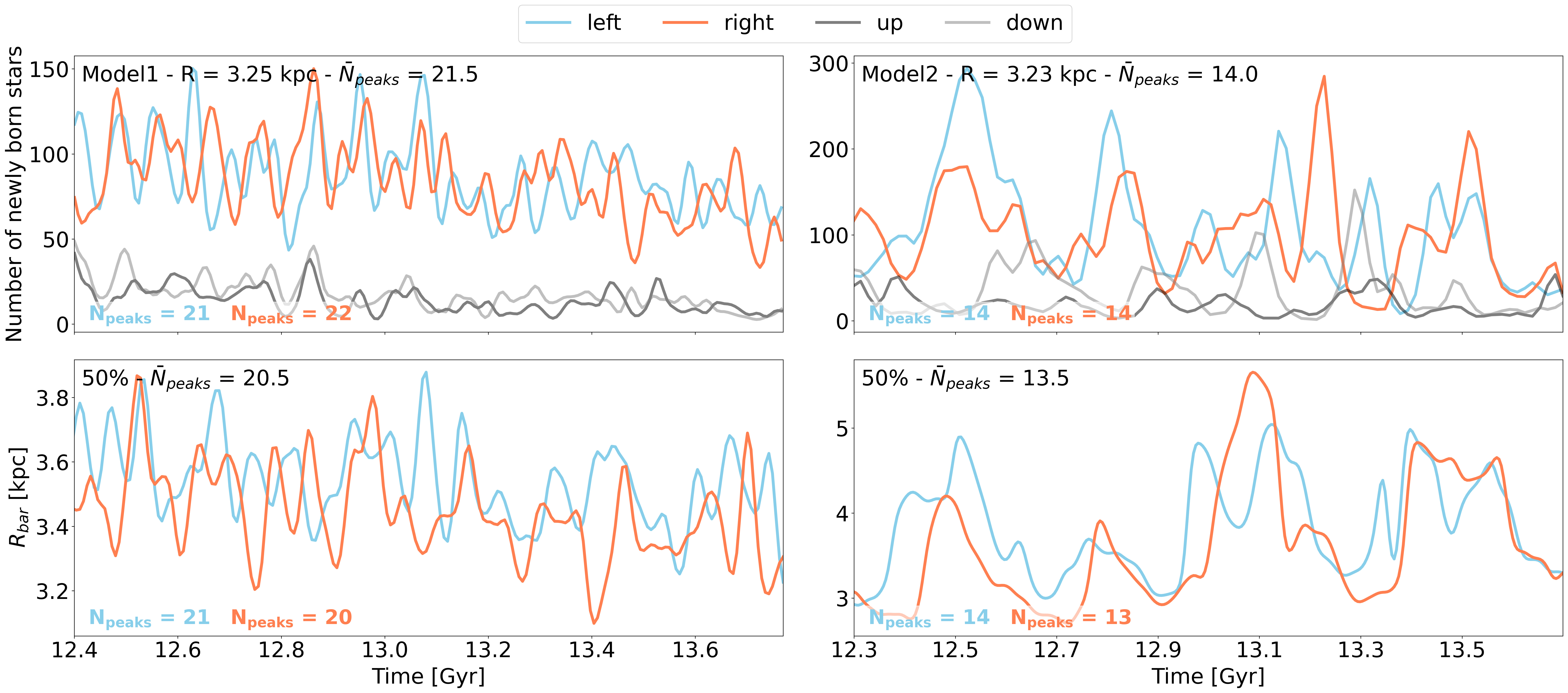}
        \caption{Comparison of the time evolution of the star formation in the four square regions used in Fig. \ref{fig:model1_sfr_diffR} (top row) to that of the bar half-length (bottom row), for Model1 and Model2 (left and right column respectively).
        Peaks in $R_{bar}$ correspond to times of bar-spiral connection. The frequency of these bar length peaks coincides with the frequency of the starbursts seen in the top panel (see number of peaks $N_{peaks}$ in each panel), suggesting their correlation with the periodic bar-spiral overlaps. Starbursts do not always happen at the same time at the two ends of the bar.}
        \label{fig:model1_sfr+rbar}
    \end{figure*}

    \begin{figure}
    \resizebox{\hsize}{!}{\includegraphics{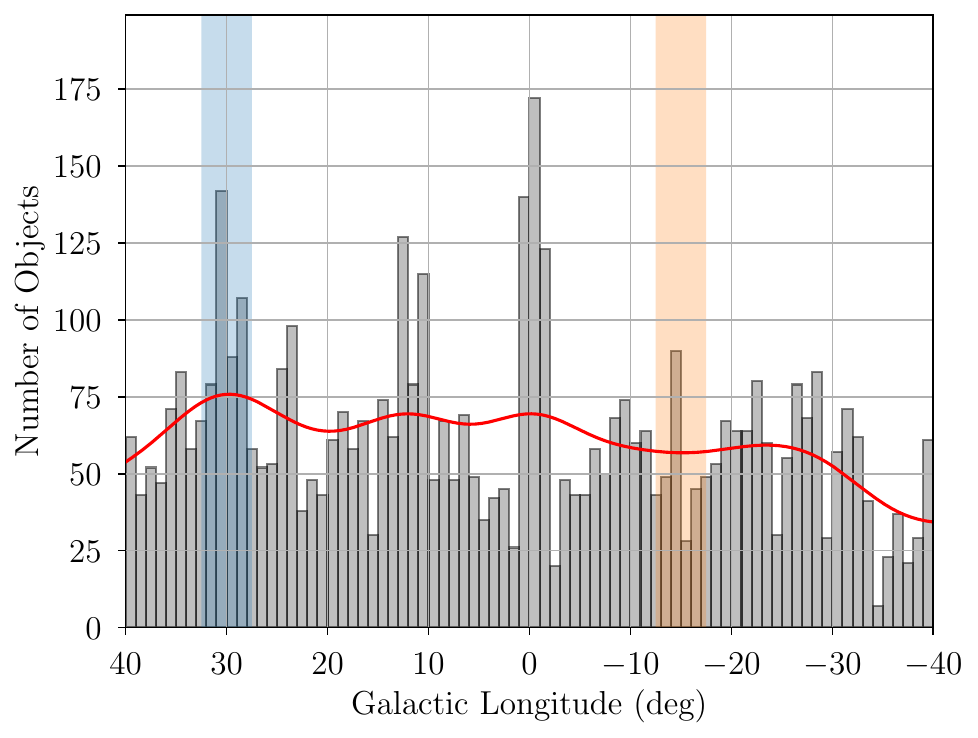}}
    \caption{Distribution of HII regions in the Milky Way as a function of galactic longitude. The red curve is a kernel density estimate of the histogram, in black, which shows data from the WISE catalog. Regions shaded in blue and orange are in the direction of the near and far side of the bar respectively. There are twice as many objects in the direction of the near end of the bar compared to that of the far end of the bar.}
    \label{fig:wise_hii}
    \end{figure}
    
    \section{Star formation at the bar radius}
    \label{sec:starformation}

    We have shown in the previous section that when a bar end and a spiral arm overlap, they create a transient overdensity and cause bursts of radial migration with frequency of the order of the bar-spiral beat frequency, typically 60 to 250 Myr, depending on the bar pattern speed. In this section, we investigate whether the star formation rates at the interface radius also follows the rythm of this periodic bar-spiral connection. We focused on four squares centered on the bar half-length radius at the four edges of the bar axes. They have the same $2\times2$ kpc$^{2}$ size as those used in the previous section (see Fig. \ref{fig:xyz_allsim+obs}), but the light blue and orange squares at each side of the bar have been combined into one square per edge, centered at $(x,y)= \pm (R_{bar},0)$. We therefore removed the distinction between leading (solid) and trailing (dashed) sides of the bar, because they did not show any difference in the star formation history. Moreover, we do not use Model3 in this section because SF is low in the region of interest, and therefore very noisy. 
    
    We report the number of stars born inside those four regions in Fig. \ref{fig:model1_sfr_diffR} at each snapshot of Model1. All curves are smoothed with a Gaussian kernel of standard deviation 10.35 Myr. The different panels are obtained by moving the squares along the same axis, therefore scanning increasing radii from top left to bottom right. The highest SF is found in the innermost bin (top left panel) along the bar axis (light blue and orange), which is actually entirely inside the bar. There is clearly much less SF along the bar minor axis. Although at this radius, the minor axis squares might also sit inside the bar envelope, bar density decreases faster vertically than horizontally, for both stars and gas, explaining lower SF activity. This does not necessarily mean that the bar ends are more efficient at forming stars than regions on the bar minor axis. Looking at the mass of gas particles in these two regions would be the only way to compare their SFE.
    
    Also notable in Fig. \ref{fig:model1_sfr_diffR} are the rapid fluctuations in SF, present along both bar axes. In the radial bin centered on the bar radius (top right), the global tendency is the same as inside the bar radius, with more extreme star formation along the bar major axis, and rapid fluctuations. However, the overall star formation is lower at these radii (bar radius) than in the previous bin (inside the bar). This is in agreement with \citet{geron2024}, who showed that for barred galaxies from the SDSS-IV MaNGA survey, star formation in the innermost regions was very high, decreased along the bar, and increased a little again just before reaching the bar radius. According to their observation, star formation in the innermost parts remains much higher than at the bar radius, which again appears consistent with our Fig. \ref{fig:model1_sfr_diffR} \citep[see alo][]{dimatteo2013, fragkoudi2018, neumann2024}. At 4.25 kpc, the difference between bar major and minor axis starts decreasing, and completely disappears from 5.25 kpc onwards as they are away from the bar ends and outside CR. Outer radii are not shown here, but we did check they followed the same trend.
    \paragraph{}To see if the fluctuations in SF at the bar radius correlate with the bar-spiral overlap, we compare them with the fluctuations in the bar length in Fig. \ref{fig:model1_sfr+rbar}. We show results for both Model1 (left column) and Model2 (right column). The top row shows the star formation history at the bar radius (top left panel is therefore the same as the top right panel of Fig. \ref{fig:model1_sfr_diffR}), while the bottom row shows the bar length measured with the $L_{cont}$ method of \citet{hilmi2020}. SF curves for Model2 are smoothed with Gaussian kernels of standard deviation 12.15 Myr.
    Model1's fast bar is characterized by a high frequency in both the bar length and star formation fluctuations. Model2 having a slower bar, fluctuations have a longer period, already revealing the importance of the bar pattern speed on star formation. As in Sec. \ref{sec:radmigr}, we counted the number of peaks in these two quantities, as indicated in the corresponding panels. SF in Model1 has 21 and 22 peaks at the left and right side of the bar respectively. This is consistent with $T_{rec}=60$ Myr found in \citet{hilmi2020}, which supports the correlation between SF and bar-spiral overlaps. For Model2 (right), four main peaks are seen both in bar length and star formation, at times 12.5 Gyr, 12.8 Gyr, 13.1 Gyr and 13.5 Gyr, i.e a peak every $\sim 325$ Myr. This corresponds to the reconnection time of the bar with the m=1 and the fast m=3 spiral modes reported in \citet{hilmi2020}. Extra higher frequency peaks are also present, suggesting the presence of other likely weaker spiral modes. Indeed, we find a total number of peaks in star formation at each end of the bar of around 13, which is consistent with the reconnection time $T_{rec} \sim 105$ Myr for the m=2, 3 and 4 spiral modes, as shown in Sec. \ref{sec:radmigr}. This supports the fact that the relation between bar-spiral coupling and star formation in the late epochs of galaxy evolution is not simulation-dependent. SF is enhanced by a factor of 3 in Model1 and 4 in Model2 when the bar and a spiral arm overlap, compared to when they disconnect.
    
    It is important to note that the number of peaks is sensitive to the smoothing width. Indeed, SF was smoothed with wider Gaussian kernels than the bar lengths (10.35 Myr vs 8.28 Myr for Model1, 12.15 Myr vs 9 Myr for Model2). Smoothing SF with the same kernels used for bar lengths, as exepected, resulted in a larger number of SF peaks. These extra peaks could be due to weak spiral modes, that are not strong enough to affect our bar length measurements, but strong enough to produce star formation.  We decided to smooth the SF with slightly larger kernels to reduce this effect. Despite this dependence on smoothing, it is clear that the galaxy with the shortest bar-spiral reconnection time produces more frequent starbursts, and the most prominent four peaks in Model2 found in star formation and bar length, as well as in migration strength, show a clear correlation with bar-spiral interaction.
    
    Finally, we also note that peaks in star formation do not necessarily happen simultaneously at the two sides of the bar (e.g., at 12.6 Gyr in Model1, or 13.2 Gyr in Model2) which must be due to a non-bisymmetric spiral structure as was indeed found in the Fourier spectral decomposition done by \citet{hilmi2020}.
    In the MW, the distribution of star formation appears asymmetric in the inner Galaxy. The WISE Catalog of Galactic HII Regions \citep{anderson2014} is a compilation of all known and candidate massive star forming regions identified by their Wide-Field Infrared Survey Explorer (WISE) infrared morphologies and correlated with radio continuum and spectroscopic ionized gas observations. \citet{armentrout2021} showed that most, if not all, HII region candidates in the WISE Catalog are bonafide HII regions, even if they lack a spectroscopic ionized gas detection. The first Galactic quadrant, which includes the near end of the bar, contains about 50\% more HII regions and HII region candidates than the fourth quadrant, which contains the far end of the bar. Considering only the ends of the bar (north: $(30 \pm 2.5) ^{\circ}$ ; south: $(-15 \pm 2.5)^{\circ}$), the asymmetry is exaggerated: there are twice as many HII regions and HII region candidates in the wedge containing the northern end of the bar compared to the wedge containing the southern end of the bar (see Fig. \ref{fig:wise_hii}).
    This can be explained by the presence of an odd spiral mode, if only one end of the bar is connected to a spiral arm, thus experiencing a burst of SF, while the other side is not (or if one end of the bar is connected to more arms than the other, in the case of a superposition of several spiral modes). A non-bisymmetric spiral structure in the MW is also consistent with \citet{khalil2024}, who recently showed that several features in the in-plane motions of disk stars, of which moving groups and phase-space ridges, could be obtained from having an axisymmetric gravitational potential, to which a bar, and an m=2 and m=3 spiral modes are added.
    
    \section{Conclusion}
    \label{sec:conclu}
    In this work, we studied the temporal evolution of the migration strength and SF of regions near the bar radius of both an N-body isolated galaxy derived from the simulation presented in \citet{vislosky2024}, as well as two cosmological simulations, one from the NIHAO-UHD project presented in \citet{buck2018_sim}, and one from the suite of \citet{martig2012}. Our results are summarized below.

    \begin{itemize}
        \item For galactic radii near the bar, stars migrate differently depending on their azimuth. Stars along the leading side of the bar's major axis seem to migrate mostly inward, while stars along the bar minor axis and along its trailing side migrate mostly outward. The difference in migration strength between the leading and trailing sides of the bar is particularly enhanced when the bar connects with a spiral.
        
        \item Near the bar radius, migration strength over $\Delta t = 100$ Myr varies regularly with time for all azimuths. Migration peaks happen when the bar length appears longer, which was recently shown to result from an overlap of the bar with spiral modes.
        
        \item The signature of bar-spiral induced migration is seen between the bar's ILR and outside its CR, beyond which other drivers take over. We speculate that this mechanism is most efficient when the bar-spiral reconnection frequency matches the stellar libration frequency.

        \item The bar-spiral overlap migration mechanism can send stars away by up to 3-4 kpc from their initial radius ($\sim R_{bar}$) in just 100 Myr. These stars reach the solar radius over longer timescales.

        \item The current SNd of both cosmological models and APOGEE data all contain a significant portion of stars coming from inside the bar radius. Those stars with a central origin are mainly on cold orbits, mostly metal-rich and quite old with their age distribution peaking around 8-10 Gyr, with a tail of younger migrators.

        \item Our cosmological model with the fastest bar (Model1) sent the most stars from the bar region to the SNd even though it has weak spirals, likely because of its late massive merger, but also potentially because its high beat frequency ($\sim$ 60 Myr vs $\sim$ 200 Myr for Model2) matches better the stellar libration frequency.

        \item The star formation rate at the bar ends fluctuates regularly, with a frequency similar to that of the bar length variations, indicating a correlation with the bar-spiral overlap.

        \item Both migration and SF bursts do not take place necessarily simultaneously at the two ends of the bar, suggesting the importance of non-bisymmetry in the spiral structure of those galaxies. Observational data from the \textit{WISE} catalog of Galactic HII sources, reveal an asymmetry in the SFR between the near and far end of the bar. We propose the presence of a non-bisymmetric spiral structure as an explanation. Similarly, resolved observations of face-on barred galaxies could provide comparisons of star formation rates from the two sides of the bar, therefore probing the symmetry of their spiral structure. 
        
    \end{itemize}

    In this work, we present bar-spiral periodic overlap as a new mechanism for efficient radial migration in disk galaxies, susceptible to move stars from orbits near bar radii to solar-like radii, mainly preserving low eccentricity. The same phenomenon also regulates the recent star formation history in our cosmological models, further highlighting the importance of bar-spiral interaction. Including this effect in subsequent analyses of both observations (e.g., upcoming 4MOST, WEAVE, SDSS-V) and numerical simulations will bring a more thorough understanding of the processes driving galactic evolution.
    
    \begin{acknowledgements}
    LM and IM thank Eugene Vasiliev for useful discussions. LM thanks the Erasmus+ program for making this joint research work possible. BR and IM acknowledge support by the Deutsche Forschungsgemeinschaft under the grant MI 2009/2-1. T.V.W. is supported by a National Science Foundation Astronomy and Astrophysics Postdoctoral Fellowship under award AST-2202340. TB’s contribution to this project was made possible by funding from the Carl Zeiss Foundation. We gratefully acknowledge the Gauss Centre for Supercomputing e.V. (www.gauss-centre.eu<http://www.gauss-centre.eu>) for funding this project by providing computing time on the GCS Supercomputer SuperMUC at Leibniz Supercomputing Centre (www.lrz.de<http://www.lrz.de>). This research was carried out on the High Performance Computing resources at New York University Abu Dhabi. GK gratefully acknowledges support from the french national research agency (ANR) funded project  MWDisc (ANR-20-CE31-0004).
    \end{acknowledgements}

    \bibliographystyle{aa} % style aa.bst
    \bibliography{biblio} % your references 

\begin{thebibliography}{159}
\expandafter\ifx\csname natexlab\endcsname\relax\def\natexlab#1{#1}\fi

\bibitem[{{Abdurro'uf} {et~al.}(2022){Abdurro'uf}, {Accetta}, {Aerts}, {Silva Aguirre}, {Ahumada}, {Ajgaonkar}, {Filiz Ak}, {Alam}, {Allende Prieto}, {Almeida}, {Anders}, {Anderson}, {Andrews}, {Anguiano}, {Aquino-Ort{\'\i}z}, {Arag{\'o}n-Salamanca}, {Argudo-Fern{\'a}ndez}, {Ata}, {Aubert}, {Avila-Reese}, {Badenes}, {Barb{\'a}}, {Barger}, {Barrera-Ballesteros}, {Beaton}, {Beers}, {Belfiore}, {Bender}, {Bernardi}, {Bershady}, {Beutler}, {Bidin}, {Bird}, {Bizyaev}, {Blanc}, {Blanton}, {Boardman}, {Bolton}, {Boquien}, {Borissova}, {Bovy}, {Brandt}, {Brown}, {Brownstein}, {Brusa}, {Buchner}, {Bundy}, {Burchett}, {Bureau}, {Burgasser}, {Cabang}, {Campbell}, {Cappellari}, {Carlberg}, {Wanderley}, {Carrera}, {Cash}, {Chen}, {Chen}, {Cherinka}, {Chiappini}, {Choi}, {Chojnowski}, {Chung}, {Clerc}, {Cohen}, {Comerford}, {Comparat}, {da Costa}, {Covey}, {Crane}, {Cruz-Gonzalez}, {Culhane}, {Cunha}, {Dai}, {Damke}, {Darling}, {Davidson}, {Davies}, {Dawson}, {De Lee}, {Diamond-Stanic}, {Cano-D{\'\i}az}, {S{\'a}nchez},
  {Donor}, {Duckworth}, {Dwelly}, {Eisenstein}, {Elsworth}, {Emsellem}, {Eracleous}, {Escoffier}, {Fan}, {Farr}, {Feng}, {Fern{\'a}ndez-Trincado}, {Feuillet}, {Filipp}, {Fillingham}, {Frinchaboy}, {Fromenteau}, {Galbany}, {Garc{\'\i}a}, {Garc{\'\i}a-Hern{\'a}ndez}, {Ge}, {Geisler}, {Gelfand}, {G{\'e}ron}, {Gibson}, {Goddy}, {Godoy-Rivera}, {Grabowski}, {Green}, {Greener}, {Grier}, {Griffith}, {Guo}, {Guy}, {Hadjara}, {Harding}, {Hasselquist}, {Hayes}, {Hearty}, {Hern{\'a}ndez}, {Hill}, {Hogg}, {Holtzman}, {Horta}, {Hsieh}, {Hsu}, {Hsu}, {Huber}, {Huertas-Company}, {Hutchinson}, {Hwang}, {Ibarra-Medel}, {Chitham}, {Ilha}, {Imig}, {Jaekle}, {Jayasinghe}, {Ji}, {Johnson}, {Jones}, {J{\"o}nsson}, {Katkov}, {Khalatyan}, {Kinemuchi}, {Kisku}, {Knapen}, {Kneib}, {Kollmeier}, {Kong}, {Kounkel}, {Kreckel}, {Krishnarao}, {Lacerna}, {Lane}, {Langgin}, {Lavender}, {Law}, {Lazarz}, {Leung}, {Leung}, {Lewis}, {Li}, {Li}, {Lian}, {Liang}, {Lin}, {Lin}, {Lin}, {Lintott}, {Long}, {Longa-Pe{\~n}a}, {L{\'o}pez-Cob{\'a}}, {Lu},
  {Lundgren}, {Luo}, {Mackereth}, {de la Macorra}, {Mahadevan}, {Majewski}, {Manchado}, {Mandeville}, {Maraston}, {Margalef-Bentabol}, {Masseron}, {Masters}, {Mathur}, {McDermid}, {Mckay}, {Merloni}, {Merrifield}, {Meszaros}, {Miglio}, {Di Mille}, {Minniti}, {Minsley}, {Monachesi}, {Moon}, {Mosser}, {Mulchaey}, {Muna}, {Mu{\~n}oz}, {Myers}, {Myers}, {Nadathur}, {Nair}, {Nandra}, {Neumann}, {Newman}, {Nidever}, {Nikakhtar}, {Nitschelm}, {O'Connell}, {Garma-Oehmichen}, {Luan Souza de Oliveira}, {Olney}, {Oravetz}, {Ortigoza-Urdaneta}, {Osorio}, {Otter}, {Pace}, {Padilla}, {Pan}, {Pan}, {Parikh}, {Parker}, {Peirani}, {Pe{\~n}a Ram{\'\i}rez}, {Penny}, {Percival}, {Perez-Fournon}, {Pinsonneault}, {Poidevin}, {Poovelil}, {Price-Whelan}, {B{\'a}rbara de Andrade Queiroz}, {Raddick}, {Ray}, {Rembold}, {Riddle}, {Riffel}, {Riffel}, {Rix}, {Robin}, {Rodr{\'\i}guez-Puebla}, {Roman-Lopes}, {Rom{\'a}n-Z{\'u}{\~n}iga}, {Rose}, {Ross}, {Rossi}, {Rubin}, {Salvato}, {S{\'a}nchez}, {S{\'a}nchez-Gallego}, {Sanderson}, {Santana
  Rojas}, {Sarceno}, {Sarmiento}, {Sayres}, {Sazonova}, {Schaefer}, {Schiavon}, {Schlegel}, {Schneider}, {Schultheis}, {Schwope}, {Serenelli}, {Serna}, {Shao}, {Shapiro}, {Sharma}, {Shen}, {Shetrone}, {Shu}, {Simon}, {Skrutskie}, {Smethurst}, {Smith}, {Sobeck}, {Spoo}, {Sprague}, {Stark}, {Stassun}, {Steinmetz}, {Stello}, {Stone-Martinez}, {Storchi-Bergmann}, {Stringfellow}, {Stutz}, {Su}, {Taghizadeh-Popp}, {Talbot}, {Tayar}, {Telles}, {Teske}, {Thakar}, {Theissen}, {Tkachenko}, {Thomas}, {Tojeiro}, {Hernandez Toledo}, {Troup}, {Trump}, {Trussler}, {Turner}, {Tuttle}, {Unda-Sanzana}, {V{\'a}zquez-Mata}, {Valentini}, {Valenzuela}, {Vargas-Gonz{\'a}lez}, {Vargas-Maga{\~n}a}, {Alfaro}, {Villanova}, {Vincenzo}, {Wake}, {Warfield}, {Washington}, {Weaver}, {Weijmans}, {Weinberg}, {Weiss}, {Westfall}, {Wild}, {Wilde}, {Wilson}, {Wilson}, {Wilson}, {Wolf}, {Wood-Vasey}, {Yan}, {Zamora}, {Zasowski}, {Zhang}, {Zhao}, {Zheng}, {Zheng}, \& {Zhu}}]{abdurro'uf2022}
{Abdurro'uf}, {Accetta}, K., {Aerts}, C., {et~al.} 2022, \apjs, 259, 35

\bibitem[{{Agertz} {et~al.}(2021){Agertz}, {Renaud}, {Feltzing}, {Read}, {Ryde}, {Andersson}, {Rey}, {Bensby}, \& {Feuillet}}]{agertz2021_vintergatan}
{Agertz}, O., {Renaud}, F., {Feltzing}, S., {et~al.} 2021, \mnras, 503, 5826

\bibitem[{{Alonso-Herrero} \& {Knapen}(2001)}]{alonso-herrero2001}
{Alonso-Herrero}, A. \& {Knapen}, J.~H. 2001, \aj, 122, 1350

\bibitem[{{Anderson} {et~al.}(2014){Anderson}, {Bania}, {Balser}, {Cunningham}, {Wenger}, {Johnstone}, \& {Armentrout}}]{anderson2014}
{Anderson}, L.~D., {Bania}, T.~M., {Balser}, D.~S., {et~al.} 2014, \apjs, 212, 1

\bibitem[{{Arifyanto} \& {Fuchs}(2006)}]{arif2006_ecc}
{Arifyanto}, M.~I. \& {Fuchs}, B. 2006, \aap, 449, 533

\bibitem[{{Armentrout} {et~al.}(2021){Armentrout}, {Anderson}, {Wenger}, {Balser}, \& {Bania}}]{armentrout2021}
{Armentrout}, W.~P., {Anderson}, L.~D., {Wenger}, T.~V., {Balser}, D.~S., \& {Bania}, T.~M. 2021, \apjs, 253, 23

\bibitem[{{Baba} {et~al.}(2013){Baba}, {Saitoh}, \& {Wada}}]{baba2013}
{Baba}, J., {Saitoh}, T.~R., \& {Wada}, K. 2013, \apj, 763, 46

\bibitem[{{Barbillon} {et~al.}(2025){Barbillon}, {Recio-Blanco}, {Poggio}, {Palicio}, {Spitoni}, {de Laverny}, \& {Cescutti}}]{barbillon2025}
{Barbillon}, M., {Recio-Blanco}, A., {Poggio}, E., {et~al.} 2025, \aap, 693, A3

\bibitem[{{Belokurov} {et~al.}(2018){Belokurov}, {Erkal}, {Evans}, {Koposov}, \& {Deason}}]{belokurov2018_gse-age}
{Belokurov}, V., {Erkal}, D., {Evans}, N.~W., {Koposov}, S.~E., \& {Deason}, A.~J. 2018, \mnras, 478, 611

\bibitem[{{Bertin} \& {Lin}(1996)}]{bertin1996}
{Bertin}, G. \& {Lin}, C.~C. 1996, {Spiral structure in galaxies a density wave theory} (Cambridge, Mass. : MIT Press)

\bibitem[{{Binney} {et~al.}(1991){Binney}, {Gerhard}, {Stark}, {Bally}, \& {Uchida}}]{binney1991}
{Binney}, J., {Gerhard}, O.~E., {Stark}, A.~A., {Bally}, J., \& {Uchida}, K.~I. 1991, \mnras, 252, 210

\bibitem[{{Binney} \& {Tremaine}(2008)}]{BT08galdyn}
{Binney}, J. \& {Tremaine}, S. 2008, {Galactic Dynamics: Second Edition} (Princeton University Press)

\bibitem[{{Bird} {et~al.}(2012){Bird}, {Kazantzidis}, \& {Weinberg}}]{bird2012}
{Bird}, J.~C., {Kazantzidis}, S., \& {Weinberg}, D.~H. 2012, \mnras, 420, 913

\bibitem[{{Bland-Hawthorn} \& {Gerhard}(2016)}]{BHG2016}
{Bland-Hawthorn}, J. \& {Gerhard}, O. 2016, \araa, 54, 529

\bibitem[{{Blanton} {et~al.}(2017){Blanton}, {Bershady}, {Abolfathi}, {Albareti}, {Allende Prieto}, {Almeida}, {Alonso-Garc{\'\i}a}, {Anders}, {Anderson}, {Andrews}, {Aquino-Ort{\'\i}z}, {Arag{\'o}n-Salamanca}, {Argudo-Fern{\'a}ndez}, {Armengaud}, {Aubourg}, {Avila-Reese}, {Badenes}, {Bailey}, {Barger}, {Barrera-Ballesteros}, {Bartosz}, {Bates}, {Baumgarten}, {Bautista}, {Beaton}, {Beers}, {Belfiore}, {Bender}, {Berlind}, {Bernardi}, {Beutler}, {Bird}, {Bizyaev}, {Blanc}, {Blomqvist}, {Bolton}, {Boquien}, {Borissova}, {van den Bosch}, {Bovy}, {Brandt}, {Brinkmann}, {Brownstein}, {Bundy}, {Burgasser}, {Burtin}, {Busca}, {Cappellari}, {Delgado Carigi}, {Carlberg}, {Carnero Rosell}, {Carrera}, {Chanover}, {Cherinka}, {Cheung}, {G{\'o}mez Maqueo Chew}, {Chiappini}, {Choi}, {Chojnowski}, {Chuang}, {Chung}, {Cirolini}, {Clerc}, {Cohen}, {Comparat}, {da Costa}, {Cousinou}, {Covey}, {Crane}, {Croft}, {Cruz-Gonzalez}, {Garrido Cuadra}, {Cunha}, {Damke}, {Darling}, {Davies}, {Dawson}, {de la Macorra}, {Dell'Agli}, {De
  Lee}, {Delubac}, {Di Mille}, {Diamond-Stanic}, {Cano-D{\'\i}az}, {Donor}, {Downes}, {Drory}, {du Mas des Bourboux}, {Duckworth}, {Dwelly}, {Dyer}, {Ebelke}, {Eigenbrot}, {Eisenstein}, {Emsellem}, {Eracleous}, {Escoffier}, {Evans}, {Fan}, {Fern{\'a}ndez-Alvar}, {Fernandez-Trincado}, {Feuillet}, {Finoguenov}, {Fleming}, {Font-Ribera}, {Fredrickson}, {Freischlad}, {Frinchaboy}, {Fuentes}, {Galbany}, {Garcia-Dias}, {Garc{\'\i}a-Hern{\'a}ndez}, {Gaulme}, {Geisler}, {Gelfand}, {Gil-Mar{\'\i}n}, {Gillespie}, {Goddard}, {Gonzalez-Perez}, {Grabowski}, {Green}, {Grier}, {Gunn}, {Guo}, {Guy}, {Hagen}, {Hahn}, {Hall}, {Harding}, {Hasselquist}, {Hawley}, {Hearty}, {Gonzalez Hern{\'a}ndez}, {Ho}, {Hogg}, {Holley-Bockelmann}, {Holtzman}, {Holzer}, {Huehnerhoff}, {Hutchinson}, {Hwang}, {Ibarra-Medel}, {da Silva Ilha}, {Ivans}, {Ivory}, {Jackson}, {Jensen}, {Johnson}, {Jones}, {J{\"o}nsson}, {Jullo}, {Kamble}, {Kinemuchi}, {Kirkby}, {Kitaura}, {Klaene}, {Knapp}, {Kneib}, {Kollmeier}, {Lacerna}, {Lane}, {Lang}, {Law},
  {Lazarz}, {Lee}, {Le Goff}, {Liang}, {Li}, {Li}, {Lian}, {Lima}, {Lin}, {Lin}, {Bertran de Lis}, {Liu}, {de Icaza Lizaola}, {Long}, {Lucatello}, {Lundgren}, {MacDonald}, {Deconto Machado}, {MacLeod}, {Mahadevan}, {Geimba Maia}, {Maiolino}, {Majewski}, {Malanushenko}, {Malanushenko}, {Manchado}, {Mao}, {Maraston}, {Marques-Chaves}, {Masseron}, {Masters}, {McBride}, {McDermid}, {McGrath}, {McGreer}, {Medina Pe{\~n}a}, {Melendez}, {Merloni}, {Merrifield}, {Meszaros}, {Meza}, {Minchev}, {Minniti}, {Miyaji}, {More}, {Mulchaey}, {M{\"u}ller-S{\'a}nchez}, {Muna}, {Munoz}, {Myers}, {Nair}, {Nandra}, {Correa do Nascimento}, {Negrete}, {Ness}, {Newman}, {Nichol}, {Nidever}, {Nitschelm}, {Ntelis}, {O'Connell}, {Oelkers}, {Oravetz}, {Oravetz}, {Pace}, {Padilla}, {Palanque-Delabrouille}, {Alonso Palicio}, {Pan}, {Parejko}, {Parikh}, {P{\^a}ris}, {Park}, {Patten}, {Peirani}, {Pellejero-Ibanez}, {Penny}, {Percival}, {Perez-Fournon}, {Petitjean}, {Pieri}, {Pinsonneault}, {Pisani}, {Poleski}, {Prada}, {Prakash}, {Queiroz},
  {Raddick}, {Raichoor}, {Barboza Rembold}, {Richstein}, {Riffel}, {Riffel}, {Rix}, {Robin}, {Rockosi}, {Rodr{\'\i}guez-Torres}, {Roman-Lopes}, {Rom{\'a}n-Z{\'u}{\~n}iga}, {Rosado}, {Ross}, {Rossi}, {Ruan}, {Ruggeri}, {Rykoff}, {Salazar-Albornoz}, {Salvato}, {S{\'a}nchez}, {Aguado}, {S{\'a}nchez-Gallego}, {Santana}, {Santiago}, {Sayres}, {Schiavon}, {da Silva Schimoia}, {Schlafly}, {Schlegel}, {Schneider}, {Schultheis}, {Schuster}, {Schwope}, {Seo}, {Shao}, {Shen}, {Shetrone}, {Shull}, {Simon}, {Skinner}, {Skrutskie}, {Slosar}, {Smith}, {Sobeck}, {Sobreira}, {Somers}, {Souto}, {Stark}, {Stassun}, {Stauffer}, {Steinmetz}, {Storchi-Bergmann}, {Streblyanska}, {Stringfellow}, {Su{\'a}rez}, {Sun}, {Suzuki}, {Szigeti}, {Taghizadeh-Popp}, {Tang}, {Tao}, {Tayar}, {Tembe}, {Teske}, {Thakar}, {Thomas}, {Thompson}, {Tinker}, {Tissera}, {Tojeiro}, {Hernandez Toledo}, {de la Torre}, {Tremonti}, {Troup}, {Valenzuela}, {Martinez Valpuesta}, {Vargas-Gonz{\'a}lez}, {Vargas-Maga{\~n}a}, {Vazquez}, {Villanova}, {Vivek}, {Vogt},
  {Wake}, {Walterbos}, {Wang}, {Weaver}, {Weijmans}, {Weinberg}, {Westfall}, {Whelan}, {Wild}, {Wilson}, {Wood-Vasey}, {Wylezalek}, {Xiao}, {Yan}, {Yang}, {Ybarra}, {Y{\`e}che}, {Zakamska}, {Zamora}, {Zarrouk}, {Zasowski}, {Zhang}, {Zhao}, {Zheng}, {Zheng}, {Zhou}, {Zhou}, {Zhu}, {Zoccali}, \& {Zou}}]{blanton2017}
{Blanton}, M.~R., {Bershady}, M.~A., {Abolfathi}, B., {et~al.} 2017, \aj, 154, 28

\bibitem[{{Blitz} \& {Spergel}(1991)}]{blitz1991}
{Blitz}, L. \& {Spergel}, D.~N. 1991, \apj, 379, 631

\bibitem[{{Boeche} {et~al.}(2013){Boeche}, {Chiappini}, {Minchev}, {Williams}, {Steinmetz}, {Sharma}, {Kordopatis}, {Bland-Hawthorn}, {Bienaym{\'e}}, {Gibson}, {Gilmore}, {Grebel}, {Helmi}, {Munari}, {Navarro}, {Parker}, {Reid}, {Seabroke}, {Siebert}, {Siviero}, {Watson}, {Wyse}, \& {Zwitter}}]{boeche2013}
{Boeche}, C., {Chiappini}, C., {Minchev}, I., {et~al.} 2013, \aap, 553, A19

\bibitem[{{Boecker} {et~al.}(2023){Boecker}, {Neumayer}, {Pillepich}, {Frankel}, {Ramesh}, {Leaman}, \& {Hernquist}}]{boecker2023_tng50-migr}
{Boecker}, A., {Neumayer}, N., {Pillepich}, A., {et~al.} 2023, \mnras, 519, 5202

\bibitem[{{Bournaud} \& {Combes}(2002)}]{bournaud2002}
{Bournaud}, F. \& {Combes}, F. 2002, \aap, 392, 83

\bibitem[{{Bournaud} \& {Combes}(2003)}]{bournaud2003}
{Bournaud}, F. \& {Combes}, F. 2003, \aap, 401, 817

\bibitem[{{Buck}(2020)}]{buck2020}
{Buck}, T. 2020, \mnras, 491, 5435

\bibitem[{{Buck} {et~al.}(2019){Buck}, {Macci{\`o}}, {Dutton}, {Obreja}, \& {Frings}}]{buck2019a}
{Buck}, T., {Macci{\`o}}, A.~V., {Dutton}, A.~A., {Obreja}, A., \& {Frings}, J. 2019, \mnras, 483, 1314

\bibitem[{{Buck} {et~al.}(2018){Buck}, {Ness}, {Macci{\`o}}, {Obreja}, \& {Dutton}}]{buck2018_sim}
{Buck}, T., {Ness}, M.~K., {Macci{\`o}}, A.~V., {Obreja}, A., \& {Dutton}, A.~A. 2018, \apj, 861, 88

\bibitem[{{Buck} {et~al.}(2023){Buck}, {Obreja}, {Ratcliffe}, {Lu}, {Minchev}, \& {Macci{\`o}}}]{buck2023}
{Buck}, T., {Obreja}, A., {Ratcliffe}, B., {et~al.} 2023, \mnras, 523, 1565

\bibitem[{{Carles} {et~al.}(2016){Carles}, {Martel}, {Ellison}, \& {Kawata}}]{carles2016}
{Carles}, C., {Martel}, H., {Ellison}, S.~L., \& {Kawata}, D. 2016, \mnras, 463, 1074

\bibitem[{{Carrillo} {et~al.}(2019){Carrillo}, {Minchev}, {Steinmetz}, {Monari}, {Laporte}, {Anders}, {Queiroz}, {Chiappini}, {Khalatyan}, {Martig}, {McMillan}, {Santiago}, \& {Youakim}}]{carrillo2019}
{Carrillo}, I., {Minchev}, I., {Steinmetz}, M., {et~al.} 2019, \mnras, 490, 797

\bibitem[{{Castro-Ginard} {et~al.}(2021){Castro-Ginard}, {McMillan}, {Luri}, {Jordi}, {Romero-G{\'o}mez}, {Cantat-Gaudin}, {Casamiquela}, {Tarricq}, {Soubiran}, \& {Anders}}]{castroginard2021}
{Castro-Ginard}, A., {McMillan}, P.~J., {Luri}, X., {et~al.} 2021, \aap, 652, A162

\bibitem[{{Cedr{\'e}s} {et~al.}(2013){Cedr{\'e}s}, {Cepa}, {Bongiovanni}, {Casta{\~n}eda}, {S{\'a}nchez-Portal}, \& {Tomita}}]{cedres2013}
{Cedr{\'e}s}, B., {Cepa}, J., {Bongiovanni}, {\'A}., {et~al.} 2013, \aap, 560, A59

\bibitem[{{Cepa} \& {Beckman}(1990)}]{cepa1990}
{Cepa}, J. \& {Beckman}, J.~E. 1990, \apj, 349, 497

\bibitem[{{Ceverino} \& {Klypin}(2007)}]{ceverino2007}
{Ceverino}, D. \& {Klypin}, A. 2007, \mnras, 379, 1155

\bibitem[{{Chiappini}(2009)}]{chiappini2009obs-migr}
{Chiappini}, C. 2009, in The Galaxy Disk in Cosmological Context, ed. J.~{Andersen}, {Nordstr{\"o}ara}, B.~{m}, \& J.~{Bland-Hawthorn}, Vol. 254, 191--196

\bibitem[{{Chiappini} {et~al.}(2003){Chiappini}, {Romano}, \& {Matteucci}}]{chiappini2003_chemevol}
{Chiappini}, C., {Romano}, D., \& {Matteucci}, F. 2003, \mnras, 339, 63

\bibitem[{{Coelho} \& {Gadotti}(2011)}]{coelho2011}
{Coelho}, P. \& {Gadotti}, D.~A. 2011, \apjl, 743, L13

\bibitem[{{Combes}(1988)}]{combes1988}
{Combes}, F. 1988, in NATO Advanced Study Institute (ASI) Series C, Vol. 232, Galactic and Extragalactic Star Formation, ed. R.~E. {Pudritz} \& M.~{Fich}, 475

\bibitem[{{Combes}(1994)}]{combes1994}
{Combes}, F. 1994, in Mass-Transfer Induced Activity in Galaxies, ed. I.~{Shlosman}, 170

\bibitem[{{Contopoulos} \& {Papayannopoulos}(1980)}]{contopoulos80a}
{Contopoulos}, G. \& {Papayannopoulos}, T. 1980, \aap, 92, 33

\bibitem[{{Di Matteo} {et~al.}(2013){Di Matteo}, {Haywood}, {Combes}, {Semelin}, \& {Snaith}}]{dimatteo2013}
{Di Matteo}, P., {Haywood}, M., {Combes}, F., {Semelin}, B., \& {Snaith}, O.~N. 2013, \aap, 553, A102

\bibitem[{{Di Matteo} {et~al.}(2019{\natexlab{a}}){Di Matteo}, {Haywood}, {Lehnert}, {Katz}, {Khoperskov}, {Snaith}, {G{\'o}mez}, \& {Robichon}}]{dimatteo2019b}
{Di Matteo}, P., {Haywood}, M., {Lehnert}, M.~D., {et~al.} 2019{\natexlab{a}}, \aap, 632, A4

\bibitem[{{Di Matteo} {et~al.}(2019{\natexlab{b}}){Di Matteo}, {Haywood}, {Lehnert}, {Katz}, {Khoperskov}, {Snaith}, {G{\'o}mez}, \& {Robichon}}]{kruijssen2020}
{Di Matteo}, P., {Haywood}, M., {Lehnert}, M.~D., {et~al.} 2019{\natexlab{b}}, \aap, 632, A4

\bibitem[{{D{\'\i}az-Garc{\'\i}a} {et~al.}(2020){D{\'\i}az-Garc{\'\i}a}, {Moyano}, {Comer{\'o}n}, {Knapen}, {Salo}, \& {Bouquin}}]{diaz-garcia2020}
{D{\'\i}az-Garc{\'\i}a}, S., {Moyano}, F.~D., {Comer{\'o}n}, S., {et~al.} 2020, \aap, 644, A38

\bibitem[{{D'Onghia} {et~al.}(2013){D'Onghia}, {Vogelsberger}, \& {Hernquist}}]{d'onghia2013}
{D'Onghia}, E., {Vogelsberger}, M., \& {Hernquist}, L. 2013, \apj, 766, 34

\bibitem[{{Eden} {et~al.}(2015){Eden}, {Moore}, {Urquhart}, {Elia}, {Plume}, {Rigby}, \& {Thompson}}]{eden2015}
{Eden}, D.~J., {Moore}, T.~J.~T., {Urquhart}, J.~S., {et~al.} 2015, \mnras, 452, 289

\bibitem[{{Eilers} {et~al.}(2019){Eilers}, {Hogg}, {Rix}, \& {Ness}}]{eilers2019}
{Eilers}, A.-C., {Hogg}, D.~W., {Rix}, H.-W., \& {Ness}, M.~K. 2019, \apj, 871, 120

\bibitem[{{Ellison} {et~al.}(2011){Ellison}, {Nair}, {Patton}, {Scudder}, {Mendel}, \& {Simard}}]{ellison2011}
{Ellison}, S.~L., {Nair}, P., {Patton}, D.~R., {et~al.} 2011, \mnras, 416, 2182

\bibitem[{{Elmegreen} \& {Elmegreen}(1986)}]{elmegreen1986}
{Elmegreen}, B.~G. \& {Elmegreen}, D.~M. 1986, \apj, 311, 554

\bibitem[{{Erwin}(2018)}]{erwin2018}
{Erwin}, P. 2018, \mnras, 474, 5372

\bibitem[{{Eskridge} {et~al.}(2000){Eskridge}, {Frogel}, {Pogge}, {Quillen}, {Davies}, {DePoy}, {Houdashelt}, {Kuchinski}, {Ram{\'\i}rez}, {Sellgren}, {Terndrup}, \& {Tiede}}]{eskridge2000}
{Eskridge}, P.~B., {Frogel}, J.~A., {Pogge}, R.~W., {et~al.} 2000, \aj, 119, 536

\bibitem[{{Feuillet} {et~al.}(2018){Feuillet}, {Bovy}, {Holtzman}, {Weinberg}, {Garc{\'\i}a-Hern{\'a}ndez}, {Hearty}, {Majewski}, {Roman-Lopes}, {Rybizki}, \& {Zamora}}]{feuillet2018}
{Feuillet}, D.~K., {Bovy}, J., {Holtzman}, J., {et~al.} 2018, \mnras, 477, 2326

\bibitem[{{Foyle} {et~al.}(2010){Foyle}, {Rix}, {Walter}, \& {Leroy}}]{foyle2010}
{Foyle}, K., {Rix}, H.~W., {Walter}, F., \& {Leroy}, A.~K. 2010, \apj, 725, 534

\bibitem[{{Fragkoudi} {et~al.}(2018){Fragkoudi}, {Di Matteo}, {Haywood}, {Schultheis}, {Khoperskov}, {G{\'o}mez}, \& {Combes}}]{fragkoudi2018}
{Fragkoudi}, F., {Di Matteo}, P., {Haywood}, M., {et~al.} 2018, \aap, 616, A180

\bibitem[{{Fraser-McKelvie} {et~al.}(2020){Fraser-McKelvie}, {Arag{\'o}n-Salamanca}, {Merrifield}, {Masters}, {Nair}, {Emsellem}, {Kraljic}, {Krishnarao}, {Andrews}, {Drory}, \& {Neumann}}]{fraser-mckelvie2020}
{Fraser-McKelvie}, A., {Arag{\'o}n-Salamanca}, A., {Merrifield}, M., {et~al.} 2020, \mnras, 495, 4158

\bibitem[{{Freeman} \& {Bland-Hawthorn}(2002)}]{freeman2002_chemtag}
{Freeman}, K. \& {Bland-Hawthorn}, J. 2002, \araa, 40, 487

\bibitem[{{Gaia Collaboration} {et~al.}(2023){Gaia Collaboration}, {Drimmel}, {Romero-G{\'o}mez}, {Chemin}, {Ramos}, {Poggio}, {Ripepi}, {Andrae}, {Blomme}, {Cantat-Gaudin}, {Castro-Ginard}, {Clementini}, {Figueras}, {Fouesneau}, {Fr{\'e}mat}, {Jardine}, {Khanna}, {Lobel}, {Marshall}, {Muraveva}, {Brown}, {Vallenari}, {Prusti}, {de Bruijne}, {Arenou}, {Babusiaux}, {Biermann}, {Creevey}, {Ducourant}, {Evans}, {Eyer}, {Guerra}, {Hutton}, {Jordi}, {Klioner}, {Lammers}, {Lindegren}, {Luri}, {Mignard}, {Panem}, {Pourbaix}, {Randich}, {Sartoretti}, {Soubiran}, {Tanga}, {Walton}, {Bailer-Jones}, {Bastian}, {Jansen}, {Katz}, {Lattanzi}, {van Leeuwen}, {Bakker}, {Cacciari}, {Casta{\~n}eda}, {De Angeli}, {Fabricius}, {Galluccio}, {Guerrier}, {Heiter}, {Masana}, {Messineo}, {Mowlavi}, {Nicolas}, {Nienartowicz}, {Pailler}, {Panuzzo}, {Riclet}, {Roux}, {Seabroke}, {Sordo}, {Th{\'e}venin}, {Gracia-Abril}, {Portell}, {Teyssier}, {Altmann}, {Audard}, {Bellas-Velidis}, {Benson}, {Berthier}, {Burgess}, {Busonero}, {Busso},
  {C{\'a}novas}, {Carry}, {Cellino}, {Cheek}, {Damerdji}, {Davidson}, {de Teodoro}, {Nu{\~n}ez Campos}, {Delchambre}, {Dell'Oro}, {Esquej}, {Fern{\'a}ndez-Hern{\'a}ndez}, {Fraile}, {Garabato}, {Garc{\'\i}a-Lario}, {Gosset}, {Haigron}, {Halbwachs}, {Hambly}, {Harrison}, {Hern{\'a}ndez}, {Hestroffer}, {Hodgkin}, {Holl}, {Jan{\ss}en}, {Jevardat de Fombelle}, {Jordan}, {Krone-Martins}, {Lanzafame}, {L{\"o}ffler}, {Marchal}, {Marrese}, {Moitinho}, {Muinonen}, {Osborne}, {Pancino}, {Pauwels}, {Recio-Blanco}, {Reyl{\'e}}, {Riello}, {Rimoldini}, {Roegiers}, {Rybizki}, {Sarro}, {Siopis}, {Smith}, {Sozzetti}, {Utrilla}, {van Leeuwen}, {Abbas}, {{\'A}brah{\'a}m}, {Abreu Aramburu}, {Aerts}, {Aguado}, {Ajaj}, {Aldea-Montero}, {Altavilla}, {{\'A}lvarez}, {Alves}, {Anders}, {Anderson}, {Anglada Varela}, {Antoja}, {Baines}, {Baker}, {Balaguer-N{\'u}{\~n}ez}, {Balbinot}, {Balog}, {Barache}, {Barbato}, {Barros}, {Barstow}, {Bartolom{\'e}}, {Bassilana}, {Bauchet}, {Becciani}, {Bellazzini}, {Berihuete}, {Bernet}, {Bertone},
  {Bianchi}, {Binnenfeld}, {Blanco-Cuaresma}, {Boch}, {Bombrun}, {Bossini}, {Bouquillon}, {Bragaglia}, {Bramante}, {Breedt}, {Bressan}, {Brouillet}, {Brugaletta}, {Bucciarelli}, {Burlacu}, {Butkevich}, {Buzzi}, {Caffau}, {Cancelliere}, {Carballo}, {Carlucci}, {Carnerero}, {Carrasco}, {Casamiquela}, {Castellani}, {Chaoul}, {Charlot}, {Chiaramida}, {Chiavassa}, {Chornay}, {Comoretto}, {Contursi}, {Cooper}, {Cornez}, {Cowell}, {Crifo}, \& {Cropper}}]{gaiacollab2023}
{Gaia Collaboration}, {Drimmel}, R., {Romero-G{\'o}mez}, M., {et~al.} 2023, \aap, 674, A37

\bibitem[{{Garc{\'\i}a P{\'e}rez} {et~al.}(2016){Garc{\'\i}a P{\'e}rez}, {Allende Prieto}, {Holtzman}, {Shetrone}, {M{\'e}sz{\'a}ros}, {Bizyaev}, {Carrera}, {Cunha}, {Garc{\'\i}a-Hern{\'a}ndez}, {Johnson}, {Majewski}, {Nidever}, {Schiavon}, {Shane}, {Smith}, {Sobeck}, {Troup}, {Zamora}, {Weinberg}, {Bovy}, {Eisenstein}, {Feuillet}, {Frinchaboy}, {Hayden}, {Hearty}, {Nguyen}, {O'Connell}, {Pinsonneault}, {Wilson}, \& {Zasowski}}]{garciaperez2016}
{Garc{\'\i}a P{\'e}rez}, A.~E., {Allende Prieto}, C., {Holtzman}, J.~A., {et~al.} 2016, \aj, 151, 144

\bibitem[{{G{\'e}ron} {et~al.}(2023){G{\'e}ron}, {Smethurst}, {Lintott}, {Kruk}, {Masters}, {Simmons}, {Mantha}, {Walmsley}, {Garma-Oehmichen}, {Drory}, \& {Lane}}]{geron2023}
{G{\'e}ron}, T., {Smethurst}, R.~J., {Lintott}, C., {et~al.} 2023, \mnras, 521, 1775

\bibitem[{{G{\'e}ron} {et~al.}(2021){G{\'e}ron}, {Smethurst}, {Lintott}, {Kruk}, {Masters}, {Simmons}, \& {Stark}}]{geron2021}
{G{\'e}ron}, T., {Smethurst}, R.~J., {Lintott}, C., {et~al.} 2021, \mnras, 507, 4389

\bibitem[{{G{\'e}ron} {et~al.}(2024){G{\'e}ron}, {Smethurst}, {Lintott}, {Masters}, {Garland}, {Mengistu}, {O'Ryan}, \& {Simmons}}]{geron2024}
{G{\'e}ron}, T., {Smethurst}, R.~J., {Lintott}, C., {et~al.} 2024, arXiv e-prints, arXiv:2405.05960

\bibitem[{{Grand} {et~al.}(2012{\natexlab{a}}){Grand}, {Kawata}, \& {Cropper}}]{grand2012a}
{Grand}, R. J.~J., {Kawata}, D., \& {Cropper}, M. 2012{\natexlab{a}}, \mnras, 426, 167

\bibitem[{{Grand} {et~al.}(2012{\natexlab{b}}){Grand}, {Kawata}, \& {Cropper}}]{grand2012b}
{Grand}, R. J.~J., {Kawata}, D., \& {Cropper}, M. 2012{\natexlab{b}}, \mnras, 421, 1529

\bibitem[{{Grenon}(1972)}]{grenon1972}
{Grenon}, M. 1972, in IAU Colloq. 17: Age des Etoiles, ed. G.~{Cayrel de Strobel} \& A.~M. {Delplace}, 55

\bibitem[{{Hackshaw} {et~al.}(2024){Hackshaw}, {Hawkins}, {Filion}, {Horta}, {Laporte}, {Carr}, \& {Price-Whelan}}]{hackshaw2024}
{Hackshaw}, Z., {Hawkins}, K., {Filion}, C., {et~al.} 2024, \apj, 977, 143

\bibitem[{{Halle} {et~al.}(2018){Halle}, {Di Matteo}, {Haywood}, \& {Combes}}]{halle2018}
{Halle}, A., {Di Matteo}, P., {Haywood}, M., \& {Combes}, F. 2018, \aap, 616, A86

\bibitem[{{Haywood} {et~al.}(2018){Haywood}, {Di Matteo}, {Lehnert}, {Snaith}, {Khoperskov}, \& {G{\'o}mez}}]{haywood2018_gse}
{Haywood}, M., {Di Matteo}, P., {Lehnert}, M.~D., {et~al.} 2018, \apj, 863, 113

\bibitem[{{Haywood} {et~al.}(2024){Haywood}, {Khoperskov}, {Cerqui}, {Di Matteo}, {Katz}, \& {Snaith}}]{haywood2024}
{Haywood}, M., {Khoperskov}, S., {Cerqui}, V., {et~al.} 2024, \aap, 690, A147

\bibitem[{{Helmi} {et~al.}(2018){Helmi}, {Babusiaux}, {Koppelman}, {Massari}, {Veljanoski}, \& {Brown}}]{helmi2018_gse}
{Helmi}, A., {Babusiaux}, C., {Koppelman}, H.~H., {et~al.} 2018, \nat, 563, 85

\bibitem[{{Hilmi} {et~al.}(2020){Hilmi}, {Minchev}, {Buck}, {Martig}, {Quillen}, {Monari}, {Famaey}, {de Jong}, {Laporte}, {Read}, {Sanders}, {Steinmetz}, \& {Wegg}}]{hilmi2020}
{Hilmi}, T., {Minchev}, I., {Buck}, T., {et~al.} 2020, \mnras, 497, 933

\bibitem[{{Holtzman} {et~al.}(2015){Holtzman}, {Shetrone}, {Johnson}, {Allende Prieto}, {Anders}, {Andrews}, {Beers}, {Bizyaev}, {Blanton}, {Bovy}, {Carrera}, {Chojnowski}, {Cunha}, {Eisenstein}, {Feuillet}, {Frinchaboy}, {Galbraith-Frew}, {Garc{\'\i}a P{\'e}rez}, {Garc{\'\i}a-Hern{\'a}ndez}, {Hasselquist}, {Hayden}, {Hearty}, {Ivans}, {Majewski}, {Martell}, {Meszaros}, {Muna}, {Nidever}, {Nguyen}, {O'Connell}, {Pan}, {Pinsonneault}, {Robin}, {Schiavon}, {Shane}, {Sobeck}, {Smith}, {Troup}, {Weinberg}, {Wilson}, {Wood-Vasey}, {Zamora}, \& {Zasowski}}]{holtzman2015}
{Holtzman}, J.~A., {Shetrone}, M., {Johnson}, J.~A., {et~al.} 2015, \aj, 150, 148

\bibitem[{{Imig} {et~al.}(2023){Imig}, {Price}, {Holtzman}, {Stone-Martinez}, {Majewski}, {Weinberg}, {Johnson}, {Allende Prieto}, {Beaton}, {Beers}, {Bizyaev}, {Blanton}, {Brownstein}, {Cunha}, {Fern{\'a}ndez-Trincado}, {Feuillet}, {Hasselquist}, {Hayes}, {J{\"o}nsson}, {Lane}, {Lian}, {M{\'e}sz{\'a}ros}, {Nidever}, {Robin}, {Shetrone}, {Smith}, \& {Wilson}}]{imig2023}
{Imig}, J., {Price}, C., {Holtzman}, J.~A., {et~al.} 2023, \apj, 954, 124

\bibitem[{{J{\"o}nsson} {et~al.}(2020){J{\"o}nsson}, {Holtzman}, {Allende Prieto}, {Cunha}, {Garc{\'\i}a-Hern{\'a}ndez}, {Hasselquist}, {Masseron}, {Osorio}, {Shetrone}, {Smith}, {Stringfellow}, {Bizyaev}, {Edvardsson}, {Majewski}, {M{\'e}sz{\'a}ros}, {Souto}, {Zamora}, {Beaton}, {Bovy}, {Donor}, {Pinsonneault}, {Poovelil}, \& {Sobeck}}]{jonsson2020}
{J{\"o}nsson}, H., {Holtzman}, J.~A., {Allende Prieto}, C., {et~al.} 2020, \aj, 160, 120

\bibitem[{{Kalnajs}(1973)}]{kalnajs1973}
{Kalnajs}, A.~J. 1973, \pasa, 2, 174

\bibitem[{{Karapetyan} {et~al.}(2018){Karapetyan}, {Hakobyan}, {Barkhudaryan}, {Mamon}, {Kunth}, {Adibekyan}, \& {Turatto}}]{karapetyan2018}
{Karapetyan}, A.~G., {Hakobyan}, A.~A., {Barkhudaryan}, L.~V., {et~al.} 2018, \mnras, 481, 566

\bibitem[{{Kawata} {et~al.}(2014){Kawata}, {Hunt}, {Grand}, {Pasetto}, \& {Cropper}}]{kawata2014_CRspiral}
{Kawata}, D., {Hunt}, J. A.~S., {Grand}, R. J.~J., {Pasetto}, S., \& {Cropper}, M. 2014, \mnras, 443, 2757

\bibitem[{{Kennicutt}(1998)}]{kennicutt1998}
{Kennicutt}, Robert~C., J. 1998, \apj, 498, 541

\bibitem[{{Khalil} {et~al.}(2024){Khalil}, {Famaey}, {Monari}, {Bernet}, {Siebert}, {Ibata}, {Thomas}, {Ramos}, {Antoja}, {Li}, {Rozier}, \& {Romero-G{\'o}mez}}]{khalil2024}
{Khalil}, Y.~R., {Famaey}, B., {Monari}, G., {et~al.} 2024, arXiv e-prints, arXiv:2411.12800

\bibitem[{{Khoperskov} {et~al.}(2020){Khoperskov}, {Di Matteo}, {Haywood}, {G{\'o}mez}, \& {Snaith}}]{Khoperskov2020_barres-sweep}
{Khoperskov}, S., {Di Matteo}, P., {Haywood}, M., {G{\'o}mez}, A., \& {Snaith}, O.~N. 2020, \aap, 638, A144

\bibitem[{{Khoperskov} \& {Gerhard}(2022)}]{khoperskov2022_sim}
{Khoperskov}, S. \& {Gerhard}, O. 2022, \aap, 663, A38

\bibitem[{{Khoperskov} {et~al.}(2018){Khoperskov}, {Haywood}, {Di Matteo}, {Lehnert}, \& {Combes}}]{khoperskov2018}
{Khoperskov}, S., {Haywood}, M., {Di Matteo}, P., {Lehnert}, M.~D., \& {Combes}, F. 2018, \aap, 609, A60

\bibitem[{{Khoperskov} {et~al.}(2021){Khoperskov}, {Haywood}, {Snaith}, {Di Matteo}, {Lehnert}, {Vasiliev}, {Naroenkov}, \& {Berczik}}]{khoperskov2021}
{Khoperskov}, S., {Haywood}, M., {Snaith}, O., {et~al.} 2021, \mnras, 501, 5176

\bibitem[{{Khoperskov} {et~al.}(2024){Khoperskov}, {Steinmetz}, {Haywood}, {van de Ven}, {Krajnovic}, {Ratcliffe}, {Minchev}, {Di Matteo}, {Kacharov}, {Marques}, {Valentini}, \& {de Jong}}]{khoperskov2024c}
{Khoperskov}, S., {Steinmetz}, M., {Haywood}, M., {et~al.} 2024, arXiv e-prints, arXiv:2411.16866

\bibitem[{{Knapen} {et~al.}(1996){Knapen}, {Beckman}, {Cepa}, \& {Nakai}}]{knapen1996}
{Knapen}, J.~H., {Beckman}, J.~E., {Cepa}, J., \& {Nakai}, N. 1996, \aap, 308, 27

\bibitem[{{Kordopatis} {et~al.}(2015){Kordopatis}, {Binney}, {Gilmore}, {Wyse}, {Belokurov}, {McMillan}, {Hatfield}, {Grebel}, {Steinmetz}, {Navarro}, {Seabroke}, {Minchev}, {Chiappini}, {Bienaym{\'e}}, {Bland-Hawthorn}, {Freeman}, {Gibson}, {Helmi}, {Munari}, {Parker}, {Reid}, {Siebert}, {Siviero}, \& {Zwitter}}]{kordo2015_obs-radmigr}
{Kordopatis}, G., {Binney}, J., {Gilmore}, G., {et~al.} 2015, \mnras, 447, 3526

\bibitem[{{Kraljic} {et~al.}(2012){Kraljic}, {Bournaud}, \& {Martig}}]{kraljic2012}
{Kraljic}, K., {Bournaud}, F., \& {Martig}, M. 2012, \apj, 757, 60

\bibitem[{{Leung} \& {Bovy}(2019)}]{leung+bovy2019}
{Leung}, H.~W. \& {Bovy}, J. 2019, \mnras, 489, 2079

\bibitem[{{Levine} {et~al.}(2006){Levine}, {Blitz}, \& {Heiles}}]{levine2006}
{Levine}, E.~S., {Blitz}, L., \& {Heiles}, C. 2006, Science, 312, 1773

\bibitem[{{Lin} \& {Shu}(1964)}]{lin1964}
{Lin}, C.~C. \& {Shu}, F.~H. 1964, \apj, 140, 646

\bibitem[{{Lin} {et~al.}(2020){Lin}, {Li}, {Du}, {Wang}, {Xiao}, {Bureau}, {Fraser-McKelvie}, {Masters}, {Lin}, {Wake}, \& {Hao}}]{lin2020}
{Lin}, L., {Li}, C., {Du}, C., {et~al.} 2020, \mnras, 499, 1406

\bibitem[{{Lord}(1987)}]{lord1987}
{Lord}, S.~D. 1987, PhD thesis, UMass Amherst

\bibitem[{{Lord} \& {Young}(1990)}]{lord1990}
{Lord}, S.~D. \& {Young}, J.~S. 1990, \apj, 356, 135

\bibitem[{{Lu} {et~al.}(2022{\natexlab{a}}){Lu}, {Minchev}, {Buck}, {Khoperskov}, {Steinmetz}, {Libeskind}, {Cescutti}, \& {Freeman}}]{lu2022b}
{Lu}, Y., {Minchev}, I., {Buck}, T., {et~al.} 2022{\natexlab{a}}, arXiv e-prints, arXiv:2212.04515

\bibitem[{{Lu} {et~al.}(2022{\natexlab{b}}){Lu}, {Ness}, {Buck}, \& {Carr}}]{lu2022a}
{Lu}, Y.~L., {Ness}, M.~K., {Buck}, T., \& {Carr}, C. 2022{\natexlab{b}}, \mnras, 512, 4697

\bibitem[{{Lucey} {et~al.}(2023){Lucey}, {Pearson}, {Hunt}, {Hawkins}, {Ness}, {Petersen}, {Price-Whelan}, \& {Weinberg}}]{lucey2023}
{Lucey}, M., {Pearson}, S., {Hunt}, J. A.~S., {et~al.} 2023, \mnras, 520, 4779

\bibitem[{{Ludlow} {et~al.}(2021){Ludlow}, {Fall}, {Schaye}, \& {Obreschkow}}]{ludlow2021}
{Ludlow}, A.~D., {Fall}, S.~M., {Schaye}, J., \& {Obreschkow}, D. 2021, \mnras, 508, 5114

\bibitem[{{Ludlow} {et~al.}(2023){Ludlow}, {Fall}, {Wilkinson}, {Schaye}, \& {Obreschkow}}]{ludlow2023}
{Ludlow}, A.~D., {Fall}, S.~M., {Wilkinson}, M.~J., {Schaye}, J., \& {Obreschkow}, D. 2023, \mnras, 525, 5614

\bibitem[{{Mackereth} \& {Bovy}(2018)}]{mackereth+bovy2018}
{Mackereth}, J.~T. \& {Bovy}, J. 2018, \pasp, 130, 114501

\bibitem[{{Maeda} {et~al.}(2020){Maeda}, {Ohta}, {Fujimoto}, {Habe}, \& {Ushio}}]{maeda2020}
{Maeda}, F., {Ohta}, K., {Fujimoto}, Y., {Habe}, A., \& {Ushio}, K. 2020, \mnras, 495, 3840

\bibitem[{{Majewski}(2017)}]{majewski2017}
{Majewski}, S. R. e.~a. 2017, \aj, 154, 94

\bibitem[{{Marinova} \& {Jogee}(2007)}]{marinova2007}
{Marinova}, I. \& {Jogee}, S. 2007, \apj, 659, 1176

\bibitem[{{Martig} {et~al.}(2012){Martig}, {Bournaud}, {Croton}, {Dekel}, \& {Teyssier}}]{martig2012}
{Martig}, M., {Bournaud}, F., {Croton}, D.~J., {Dekel}, A., \& {Teyssier}, R. 2012, \apj, 756, 26

\bibitem[{{Martig} {et~al.}(2009){Martig}, {Bournaud}, {Teyssier}, \& {Dekel}}]{martig2009}
{Martig}, M., {Bournaud}, F., {Teyssier}, R., \& {Dekel}, A. 2009, \apj, 707, 250

\bibitem[{{Martig} {et~al.}(2014{\natexlab{a}}){Martig}, {Minchev}, \& {Flynn}}]{martig2014a}
{Martig}, M., {Minchev}, I., \& {Flynn}, C. 2014{\natexlab{a}}, \mnras, 442, 2474

\bibitem[{{Martig} {et~al.}(2014{\natexlab{b}}){Martig}, {Minchev}, \& {Flynn}}]{martig2014b}
{Martig}, M., {Minchev}, I., \& {Flynn}, C. 2014{\natexlab{b}}, \mnras, 443, 2452

\bibitem[{{Men{\'e}ndez-Delmestre} {et~al.}(2007){Men{\'e}ndez-Delmestre}, {Sheth}, {Schinnerer}, {Jarrett}, \& {Scoville}}]{menendez-delmestre2007}
{Men{\'e}ndez-Delmestre}, K., {Sheth}, K., {Schinnerer}, E., {Jarrett}, T.~H., \& {Scoville}, N.~Z. 2007, \apj, 657, 790

\bibitem[{{Minchev} {et~al.}(2018){Minchev}, {Anders}, {Recio-Blanco}, {Chiappini}, {de Laverny}, {Queiroz}, {Steinmetz}, {Adibekyan}, {Carrillo}, {Cescutti}, {Guiglion}, {Hayden}, {de Jong}, {Kordopatis}, {Majewski}, {Martig}, \& {Santiago}}]{minchev2018_rb}
{Minchev}, I., {Anders}, F., {Recio-Blanco}, A., {et~al.} 2018, \mnras, 481, 1645

\bibitem[{{Minchev} {et~al.}(2013){Minchev}, {Chiappini}, \& {Martig}}]{minchev2013}
{Minchev}, I., {Chiappini}, C., \& {Martig}, M. 2013, \aap, 558, A9

\bibitem[{{Minchev} \& {Famaey}(2010)}]{minchev2010}
{Minchev}, I. \& {Famaey}, B. 2010, \apj, 722, 112

\bibitem[{{Minchev} {et~al.}(2011){Minchev}, {Famaey}, {Combes}, {Di Matteo}, {Mouhcine}, \& {Wozniak}}]{minchev2011_res-overlap}
{Minchev}, I., {Famaey}, B., {Combes}, F., {et~al.} 2011, \aap, 527, A147

\bibitem[{{Minchev} {et~al.}(2012){Minchev}, {Famaey}, {Quillen}, {Di Matteo}, {Combes}, {Vlaji{\'c}}, {Erwin}, \& {Bland-Hawthorn}}]{minchev2012a}
{Minchev}, I., {Famaey}, B., {Quillen}, A.~C., {et~al.} 2012, \aap, 548, A126

\bibitem[{{Minchev} \& {Quillen}(2006)}]{minchev2006}
{Minchev}, I. \& {Quillen}, A.~C. 2006, \mnras, 368, 623

\bibitem[{{Moore} {et~al.}(2012){Moore}, {Urquhart}, {Morgan}, \& {Thompson}}]{moore2012}
{Moore}, T.~J.~T., {Urquhart}, J.~S., {Morgan}, L.~K., \& {Thompson}, M.~A. 2012, \mnras, 426, 701

\bibitem[{{Nakada} {et~al.}(1991){Nakada}, {Onaka}, {Yamamura}, {Deguchi}, {Hashimoto}, {Izumiura}, \& {Sekiguchi}}]{nakada1991}
{Nakada}, Y., {Onaka}, T., {Yamamura}, I., {et~al.} 1991, \nat, 353, 140

\bibitem[{{Nepal} {et~al.}(2024){Nepal}, {Chiappini}, {Guiglion}, {Steinmetz}, {P{\'e}rez-Villegas}, {Queiroz}, {Miglio}, {Dohme}, \& {Khalatyan}}]{nepal2024}
{Nepal}, S., {Chiappini}, C., {Guiglion}, G., {et~al.} 2024, \aap, 681, L8

\bibitem[{{Neumann} {et~al.}(2024){Neumann}, {Thomas}, {Maraston}, {Gleis}, {Mao}, {Schinnerer}, \& {Stuber}}]{neumann2024}
{Neumann}, J., {Thomas}, D., {Maraston}, C., {et~al.} 2024, \mnras, 534, 2438

\bibitem[{{Okalidis} {et~al.}(2022){Okalidis}, {Grand}, {Yates}, \& {Springel}}]{okalidis2022_auriga-migr}
{Okalidis}, P., {Grand}, R. J.~J., {Yates}, R.~M., \& {Springel}, V. 2022, \mnras, 514, 5085

\bibitem[{{Peters}(1975)}]{peters1975}
{Peters}, III, W.~L. 1975, \apj, 195, 617

\bibitem[{Petersen {et~al.}(2019)Petersen, Weinberg, \& Katz}]{petersen2019}
Petersen, M.~S., Weinberg, M.~D., \& Katz, N. 2019, Monthly Notices of the Royal Astronomical Society, 490, 3616–3632

\bibitem[{{Poggio} {et~al.}(2021){Poggio}, {Drimmel}, {Cantat-Gaudin}, {Ramos}, {Ripepi}, {Zari}, {Andrae}, {Blomme}, {Chemin}, {Clementini}, {Figueras}, {Fouesneau}, {Fr{\'e}mat}, {Lobel}, {Marshall}, {Muraveva}, \& {Romero-G{\'o}mez}}]{poggio2021}
{Poggio}, E., {Drimmel}, R., {Cantat-Gaudin}, T., {et~al.} 2021, \aap, 651, A104

\bibitem[{{Poggio} {et~al.}(2022){Poggio}, {Recio-Blanco}, {Palicio}, {Re Fiorentin}, {de Laverny}, {Drimmel}, {Kordopatis}, {Lattanzi}, {Schultheis}, {Spagna}, \& {Spitoni}}]{poggio2022}
{Poggio}, E., {Recio-Blanco}, A., {Palicio}, P.~A., {et~al.} 2022, \aap, 666, L4

\bibitem[{{Querejeta} {et~al.}(2024){Querejeta}, {Leroy}, {Meidt}, {Schinnerer}, {Belfiore}, {Emsellem}, {Klessen}, {Sun}, {Sormani}, {Be{\v{s}}li{\'c}}, {Cao}, {Chevance}, {Colombo}, {Dale}, {Garc{\'\i}a-Burillo}, {Glover}, {Grasha}, {Groves}, {Koch}, {Neumann}, {Pan}, {Pessa}, {Pety}, {Pinna}, {Ramambason}, {Razza}, {Romanelli}, {Rosolowsky}, {Ruiz-Garc{\'\i}a}, {S{\'a}nchez-Bl{\'a}zquez}, {Smith}, {Stuber}, {Ubeda}, {Usero}, \& {Williams}}]{querejeta2024}
{Querejeta}, M., {Leroy}, A.~K., {Meidt}, S.~E., {et~al.} 2024, \aap, 687, A293

\bibitem[{{Querejeta} {et~al.}(2021){Querejeta}, {Schinnerer}, {Meidt}, {Sun}, {Leroy}, {Emsellem}, {Klessen}, {Mu{\~n}oz-Mateos}, {Salo}, {Laurikainen}, {Be{\v{s}}li{\'c}}, {Blanc}, {Chevance}, {Dale}, {Eibensteiner}, {Faesi}, {Garc{\'\i}a-Rodr{\'\i}guez}, {Glover}, {Grasha}, {Henshaw}, {Herrera}, {Hughes}, {Kreckel}, {Kruijssen}, {Liu}, {Murphy}, {Pan}, {Pety}, {Razza}, {Rosolowsky}, {Saito}, {Schruba}, {Usero}, {Watkins}, \& {Williams}}]{querejeta2021}
{Querejeta}, M., {Schinnerer}, E., {Meidt}, S., {et~al.} 2021, \aap, 656, A133

\bibitem[{{Quillen} {et~al.}(2011){Quillen}, {Dougherty}, {Bagley}, {Minchev}, \& {Comparetta}}]{quillen2011}
{Quillen}, A.~C., {Dougherty}, J., {Bagley}, M.~B., {Minchev}, I., \& {Comparetta}, J. 2011, \mnras, 417, 762

\bibitem[{{Quillen} {et~al.}(2009){Quillen}, {Minchev}, {Bland-Hawthorn}, \& {Haywood}}]{quillen2009_mergermigr}
{Quillen}, A.~C., {Minchev}, I., {Bland-Hawthorn}, J., \& {Haywood}, M. 2009, \mnras, 397, 1599

\bibitem[{{Ragan} {et~al.}(2018){Ragan}, {Moore}, {Eden}, {Hoare}, {Urquhart}, {Elia}, \& {Molinari}}]{ragan2018}
{Ragan}, S.~E., {Moore}, T.~J.~T., {Eden}, D.~J., {et~al.} 2018, \mnras, 479, 2361

\bibitem[{{Ratcliffe} {et~al.}(2024){Ratcliffe}, {Khoperskov}, {Minchev}, {Lee}, {Buck}, {Marques}, {Lu}, \& {Steinmetz}}]{ratcliffe2024c}
{Ratcliffe}, B., {Khoperskov}, S., {Minchev}, I., {et~al.} 2024, arXiv e-prints, arXiv:2410.17326

\bibitem[{{Ratcliffe} {et~al.}(2023){Ratcliffe}, {Minchev}, {Anders}, {Khoperskov}, {Guiglion}, {Buck}, {Cunha}, {Queiroz}, {Nitschelm}, {Meszaros}, {Steinmetz}, {de Jong}, {Nepal}, {Lane}, \& {Sobeck}}]{ratcliffe2023_rb}
{Ratcliffe}, B., {Minchev}, I., {Anders}, F., {et~al.} 2023, arXiv e-prints, arXiv:2305.13378

\bibitem[{{Rebolledo} {et~al.}(2012){Rebolledo}, {Wong}, {Leroy}, {Koda}, \& {Donovan Meyer}}]{rebolledo2012}
{Rebolledo}, D., {Wong}, T., {Leroy}, A., {Koda}, J., \& {Donovan Meyer}, J. 2012, \apj, 757, 155

\bibitem[{{Renaud} {et~al.}(2024){Renaud}, {Ratcliffe}, {Minchev}, {Haywood}, {Di Matteo}, {Agertz}, \& {Romeo}}]{renaud2024}
{Renaud}, F., {Ratcliffe}, B., {Minchev}, I., {et~al.} 2024, arXiv e-prints, arXiv:2409.10598

\bibitem[{{Reynaud} \& {Downes}(1998)}]{reynaud1998}
{Reynaud}, D. \& {Downes}, D. 1998, \aap, 337, 671

\bibitem[{{Roca-F{\`a}brega} {et~al.}(2013){Roca-F{\`a}brega}, {Valenzuela}, {Figueras}, {Romero-G{\'o}mez}, {Vel{\'a}zquez}, {Antoja}, \& {Pichardo}}]{roca-fabrega2013}
{Roca-F{\`a}brega}, S., {Valenzuela}, O., {Figueras}, F., {et~al.} 2013, \mnras, 432, 2878

\bibitem[{{Ro{\v{s}}kar} {et~al.}(2008){Ro{\v{s}}kar}, {Debattista}, {Quinn}, {Stinson}, \& {Wadsley}}]{Roskar2008_migration}
{Ro{\v{s}}kar}, R., {Debattista}, V.~P., {Quinn}, T.~R., {Stinson}, G.~S., \& {Wadsley}, J. 2008, \apjl, 684, L79

\bibitem[{{Sakamoto} {et~al.}(1999){Sakamoto}, {Okumura}, {Ishizuki}, \& {Scoville}}]{sakamoto1999}
{Sakamoto}, K., {Okumura}, S.~K., {Ishizuki}, S., \& {Scoville}, N.~Z. 1999, \apj, 525, 691

\bibitem[{{Sanders} {et~al.}(2024){Sanders}, {Kawata}, {Matsunaga}, {Sormani}, {Smith}, {Minniti}, \& {Gerhard}}]{sanders2024}
{Sanders}, J.~L., {Kawata}, D., {Matsunaga}, N., {et~al.} 2024, \mnras, 530, 2972

\bibitem[{{Sch{\"o}nrich} \& {Binney}(2009)}]{schoenrich2009_chem-migr}
{Sch{\"o}nrich}, R. \& {Binney}, J. 2009, \mnras, 396, 203

\bibitem[{{Seigar} \& {James}(2002)}]{seigar2002}
{Seigar}, M.~S. \& {James}, P.~A. 2002, \mnras, 337, 1113

\bibitem[{{Sellwood}(2013)}]{sellwood2013}
{Sellwood}, J.~A. 2013, \apjl, 769, L24

\bibitem[{{Sellwood} \& {Binney}(2002)}]{SB02}
{Sellwood}, J.~A. \& {Binney}, J.~J. 2002, \mnras, 336, 785

\bibitem[{{Sestito} {et~al.}(2021){Sestito}, {Buck}, {Starkenburg}, {Martin}, {Navarro}, {Venn}, {Obreja}, {Jablonka}, \& {Macci{\`o}}}]{sestito2021}
{Sestito}, F., {Buck}, T., {Starkenburg}, E., {et~al.} 2021, \mnras, 500, 3750

\bibitem[{{Sheth} {et~al.}(2008){Sheth}, {Elmegreen}, {Elmegreen}, {Capak}, {Abraham}, {Athanassoula}, {Ellis}, {Mobasher}, {Salvato}, {Schinnerer}, {Scoville}, {Spalsbury}, {Strubbe}, {Carollo}, {Rich}, \& {West}}]{sheth2008}
{Sheth}, K., {Elmegreen}, D.~M., {Elmegreen}, B.~G., {et~al.} 2008, \apj, 675, 1141

\bibitem[{{Spitoni} {et~al.}(2019){Spitoni}, {Cescutti}, {Minchev}, {Matteucci}, {Silva Aguirre}, {Martig}, {Bono}, \& {Chiappini}}]{spitoni2019}
{Spitoni}, E., {Cescutti}, G., {Minchev}, I., {et~al.} 2019, \aap, 628, A38

\bibitem[{{Stanek} {et~al.}(1997){Stanek}, {Udalski}, {Szyma{\'N}ski}, {Ka{\L}u{\.Z}ny}, {Kubiak}, {Mateo}, \& {Krzemi{\'N}ski}}]{stanek1997}
{Stanek}, K.~Z., {Udalski}, A., {Szyma{\'N}ski}, M., {et~al.} 1997, \apj, 477, 163

\bibitem[{{Stinson} {et~al.}(2006){Stinson}, {Seth}, {Katz}, {Wadsley}, {Governato}, \& {Quinn}}]{stinson2006_SF+feedback}
{Stinson}, G., {Seth}, A., {Katz}, N., {et~al.} 2006, \mnras, 373, 1074

\bibitem[{{Stinson} {et~al.}(2013){Stinson}, {Brook}, {Macci{\`o}}, {Wadsley}, {Quinn}, \& {Couchman}}]{stinson2013_feedback}
{Stinson}, G.~S., {Brook}, C., {Macci{\`o}}, A.~V., {et~al.} 2013, \mnras, 428, 129

\bibitem[{{Stone-Martinez} {et~al.}(2024){Stone-Martinez}, {Holtzman}, {Imig}, {Nitschelm}, {Stassun}, \& {Brownstein}}]{stone-martinez2024}
{Stone-Martinez}, A., {Holtzman}, J.~A., {Imig}, J., {et~al.} 2024, \aj, 167, 73

\bibitem[{{Sun} {et~al.}(2024){Sun}, {Calzetti}, \& {Battisti}}]{sun2024}
{Sun}, B., {Calzetti}, D., \& {Battisti}, A.~J. 2024, arXiv e-prints, arXiv:2407.07248

\bibitem[{{Sygnet} {et~al.}(1988){Sygnet}, {Tagger}, {Athanassoula}, \& {Pellat}}]{sygnet1988}
{Sygnet}, J.~F., {Tagger}, M., {Athanassoula}, E., \& {Pellat}, R. 1988, \mnras, 232, 733

\bibitem[{{Tagger} {et~al.}(1987){Tagger}, {Sygnet}, {Athanassoula}, \& {Pellat}}]{tagger1987}
{Tagger}, M., {Sygnet}, J.~F., {Athanassoula}, E., \& {Pellat}, R. 1987, \apjl, 318, L43

\bibitem[{{Teyssier}(2002)}]{teyssier2002}
{Teyssier}, R. 2002, \aap, 385, 337

\bibitem[{{Toomre} \& {Kalnajs}(1991)}]{toomre1991}
{Toomre}, A. \& {Kalnajs}, A.~J. 1991, in Dynamics of Disc Galaxies, ed. B.~{Sundelius}, 341

\bibitem[{{Vera-Ciro} {et~al.}(2014){Vera-Ciro}, {D'Onghia}, {Navarro}, \& {Abadi}}]{vera-ciro14}
{Vera-Ciro}, C., {D'Onghia}, E., {Navarro}, J., \& {Abadi}, M. 2014, \apj, 794, 173

\bibitem[{{Vincenzo} \& {Kobayashi}(2020)}]{vincenzo2020}
{Vincenzo}, F. \& {Kobayashi}, C. 2020, \mnras, 496, 80

\bibitem[{{Vislosky} {et~al.}(2024){Vislosky}, {Minchev}, {Khoperskov}, {Martig}, {Buck}, {Hilmi}, {Ratcliffe}, {Bland-Hawthorn}, {Quillen}, {Steinmetz}, \& {de Jong}}]{vislosky2024}
{Vislosky}, E., {Minchev}, I., {Khoperskov}, S., {et~al.} 2024, \mnras, 528, 3576

\bibitem[{{Vogel} {et~al.}(1988){Vogel}, {Kulkarni}, \& {Scoville}}]{vogel1988}
{Vogel}, S.~N., {Kulkarni}, S.~R., \& {Scoville}, N.~Z. 1988, \nat, 334, 402

\bibitem[{{Wadsley} {et~al.}(2017){Wadsley}, {Keller}, \& {Quinn}}]{wadsley2017_gasoline2}
{Wadsley}, J.~W., {Keller}, B.~W., \& {Quinn}, T.~R. 2017, \mnras, 471, 2357

\bibitem[{{Wang} {et~al.}(2023){Wang}, {Carrillo}, {Ness}, \& {Buck}}]{wang2023}
{Wang}, K., {Carrillo}, A., {Ness}, M., \& {Buck}, T. 2023, in American Astronomical Society Meeting Abstracts, Vol. 241, American Astronomical Society Meeting Abstracts, 208.02

\bibitem[{{Wang} {et~al.}(2015){Wang}, {Dutton}, {Stinson}, {Macci{\`o}}, {Penzo}, {Kang}, {Keller}, \& {Wadsley}}]{wang2015_nihao}
{Wang}, L., {Dutton}, A.~A., {Stinson}, G.~S., {et~al.} 2015, \mnras, 454, 83

\bibitem[{{Watanabe} {et~al.}(2011){Watanabe}, {Sorai}, {Kuno}, \& {Habe}}]{watanabe2011}
{Watanabe}, Y., {Sorai}, K., {Kuno}, N., \& {Habe}, A. 2011, \mnras, 411, 1409

\bibitem[{{Weiland} {et~al.}(1994){Weiland}, {Arendt}, {Berriman}, {Dwek}, {Freudenreich}, {Hauser}, {Kelsall}, {Lisse}, {Mitra}, {Moseley}, {Odegard}, {Silverberg}, {Sodroski}, {Spiesman}, \& {Stemwedel}}]{weiland1994}
{Weiland}, J.~L., {Arendt}, R.~G., {Berriman}, G.~B., {et~al.} 1994, \apjl, 425, L81

\bibitem[{Wilkinson {et~al.}(2023)Wilkinson, Ludlow, Lagos, Fall, Schaye, \& Obreschkow}]{wilkinson2023}
Wilkinson, M.~J., Ludlow, A.~D., Lagos, C. d.~P., {et~al.} 2023, Monthly Notices of the Royal Astronomical Society, 519, 5942–5961

\bibitem[{{Zhang} {et~al.}(2024){Zhang}, {Belokurov}, {Evans}, {Kane}, \& {Sanders}}]{zhang2024a}
{Zhang}, H., {Belokurov}, V., {Evans}, N.~W., {Kane}, S.~G., \& {Sanders}, J.~L. 2024, \mnras [\eprint[arXiv]{2406.06678}]

\bibitem[{{Zurita} {et~al.}(2004){Zurita}, {Rela{\~n}o}, {Beckman}, \& {Knapen}}]{zurita2004}
{Zurita}, A., {Rela{\~n}o}, M., {Beckman}, J.~E., \& {Knapen}, J.~H. 2004, \aap, 413, 73

\end{thebibliography}

\begin{appendix}
    \section{Observational biases}
    In order to properly compare APOGEE DR17 data with the simulated galaxies, we biased the last snapshot of the models with APOGEE selection effects and uncertainties. This was only applied in section \ref{subsec:impactSNd} when comparing the last snapshot of the models to the observations.

    \subsection{Selection effects}
    \label{selection_effects}
\begin{figure*}
    \centering
        \includegraphics[width=14cm]{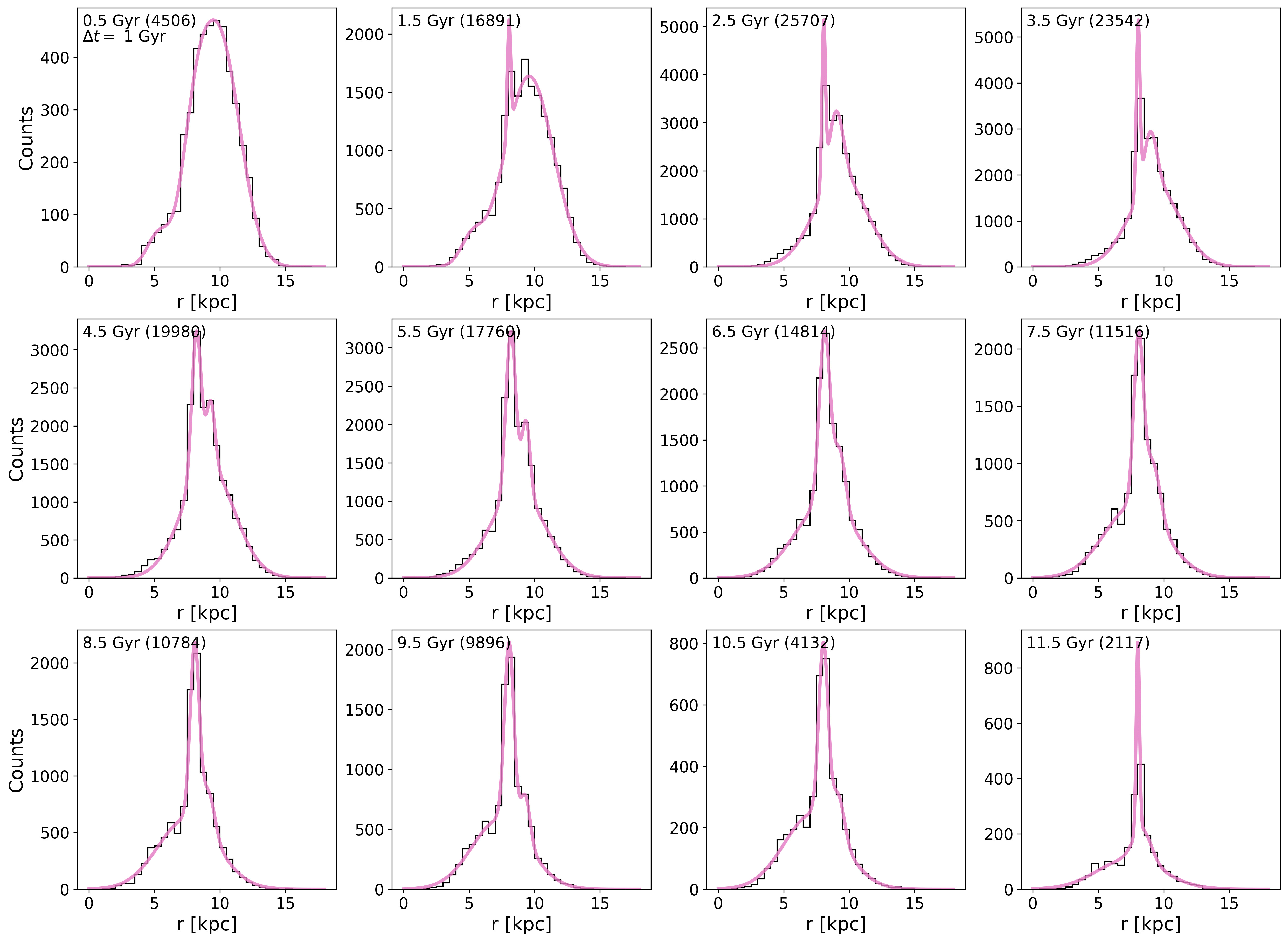}
        \caption{Radial distributions of different monoage bins in APOGEE DR17 sample. The pink curve is the best fit to the data.}
        \label{apdx:selfunc}
    \end{figure*}

    The two most important parameters of our analysis were radial positions and ages. Therefore, we tried to reproduce the radial distributions of monoage (1 Gyr wide) distributions seen in the APOGEE sample. The oldest stars not being very numerous, we grouped all stars older than 11 Gyr into one bin. The first step was to fit these distributions, histrogrammed in 0.5 kpc-wide bins, with a sum of 1, 2 or 3 Gaussians, each time choosing the model minimising the Corrected Akaike Information Criterion. This criterion allows to choose the simplest model (smallest amount of parameters) that best fits the data (smallest residual). The result of this first step is shown in Fig. \ref{apdx:selfunc}.
    The next step was to sample the simulations so that they had similar monoage radial distributions. To achieve this, we implemented a rejection sampling algorithm. Rejection sampling starts from a large original sample, and samples it to a more restricted target distribution. Here the original samples were the simulations' monoage radial distributions, binned in 0.5 kpc-wide cells, while the target distributions were the selection functions fitted from the data (Fig. \ref{apdx:selfunc}). To avoid biases towards the most populated age bins, each age bin was given the same number of stars as counted in the APOGEE sample. For each monoage populations, the same rejection sampling procedure was applied as follows. 
    \begin{enumerate}
        \item Randomly pick a star in the simulation.
        \item Find the radial bin it belongs to.
        \item Pick a random number below the total number of stars found in this radial bin. If this number is below the selection function calculated at the radius chosen, then the star is accepted. Otherwise, it is rejected.
        \item Repeat until there are as many accepted stars as observed in the corresponding APOGEE bin.
    \end{enumerate}
    
\begin{figure*}
    \centering
        \includegraphics[width=14cm]{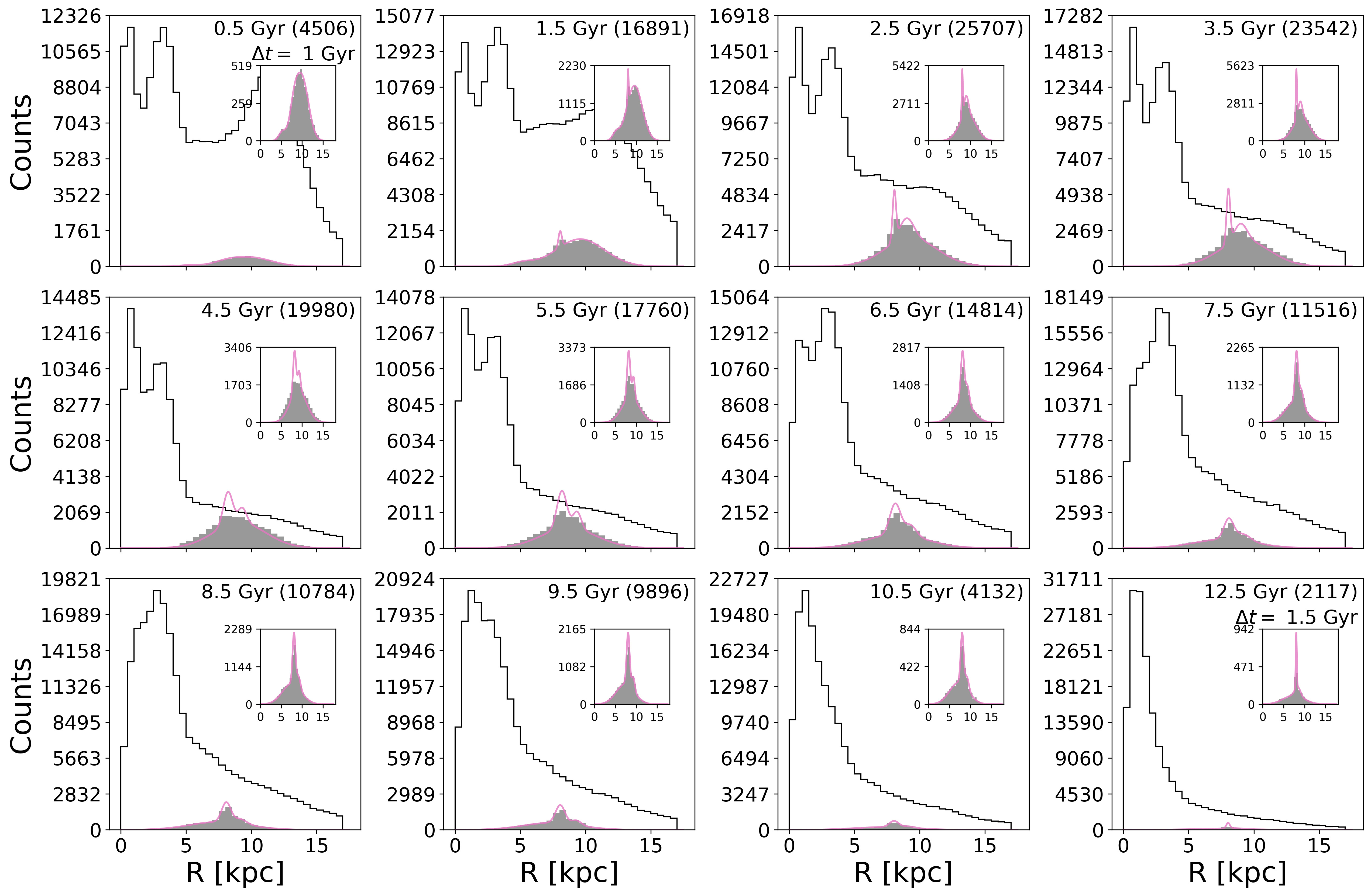}
        \caption{Final stars (filled gray) chosen from the total simulation (black line). They were selected so that each monoage populations of the sample are similarly radially distributed as in the APOGEE sample (pink lines).}
        \label{apdx:model1_rejsamp}
    \end{figure*}

    The result of this procedure is shown for Model1 in Fig. \ref{apdx:model1_rejsamp}.

    We tried constraining the sample to azimuthal ranges consistent with the position of the Sun with respect to the bar in the Milky Way, but since the Galaxy is already well phase-mixed, this did not significantly impact our results. We therefore kept the azimuthally-averaged method described above.

    \subsection{Uncertainties}
\begin{figure*}
    \centering
        \includegraphics[width=14cm]{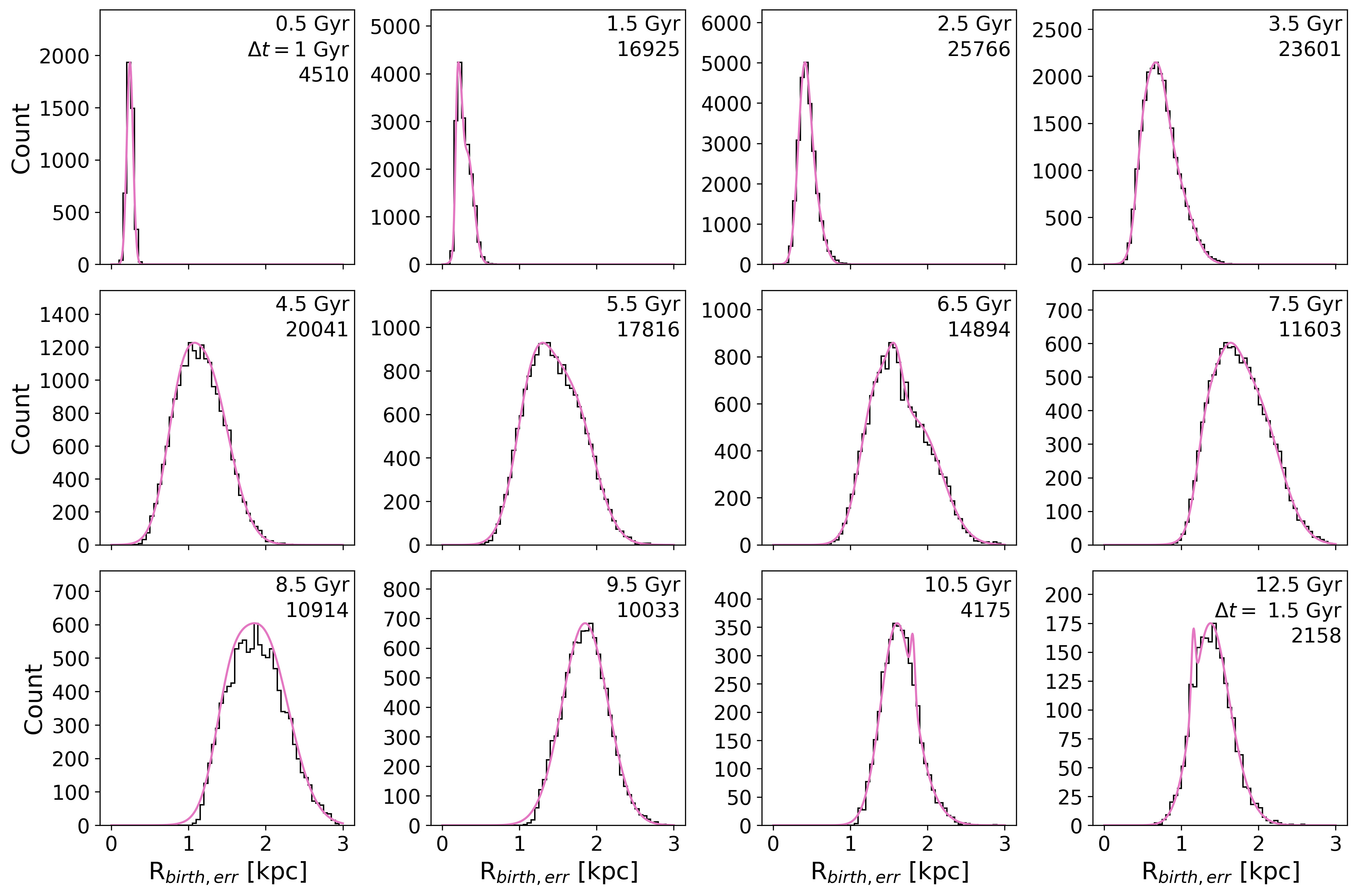}
        \caption{Birth radius error distributions of different monoage bins in the APOGEE DR17 sample. The pink curves are the best fit to the data.}
        \label{apdx:obs_Rb-err}
    \end{figure*}

\begin{figure*}
    \centering
        \includegraphics[width=14cm]{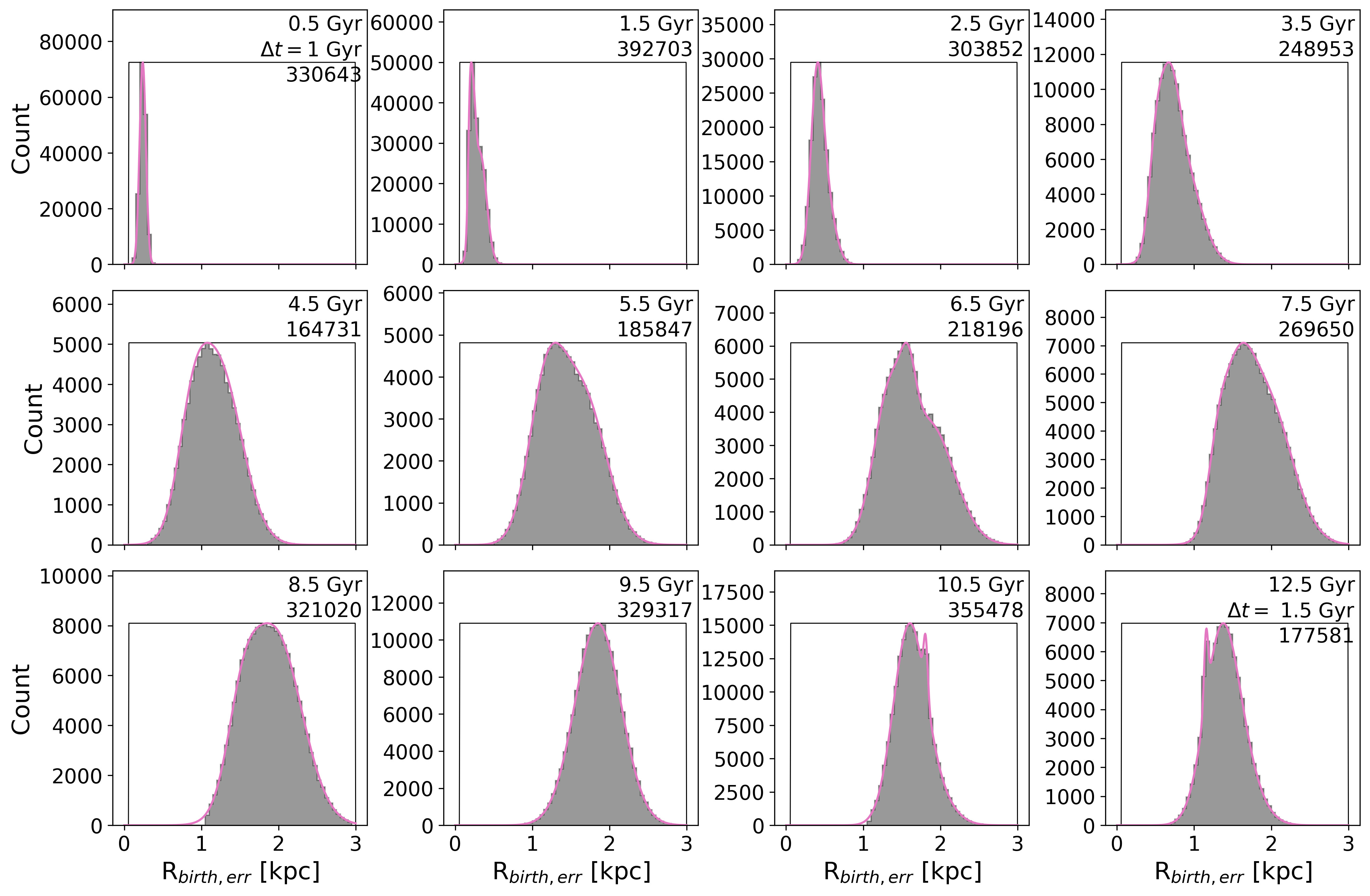}
        \caption{Modeled error on birth radius for different monoage bins in Model1. The black curve is a default arbitrary uniform distribution of errors. The pink curve is the APOGEE birth radii error distributions. The filled gray histogram is the error distribution drawn from the uniform distribution (black) so that it fits the data (pink).}
        \label{apdx:model1_rb-err}
    \end{figure*}

    We implemented Gaussian errors in the simulations for $r$, $z$, $v_r$, $v_{phi}$, $v_z$, $age$ and [Fe/H] with standard deviations reported in Table\ref{table:errors}. 
    The errors on birth radius depending largely on age, we modeled their uncertainties on an age-by-age basis, similarly as when accounting for the selection effects. 
    We fitted the distribution of APOGEE birth radius errors for monoage populations (1 Gyr wide) with 1, 2 or 3 gaussians, keeping the model that minimized the Corrected Akaike Information Criterion. 
    In Fig. \ref{apdx:obs_Rb-err}, histograms of the error distributions for different monoage bins of the APOGEE simple are shown as the black curves. 
    The pink curves are the fitted models. We implemented these uncertainties using again a rejection sampling method. 
    For each age bin, the original sample was artificially chosen as uniform between the minimal and maximal errors ($E_{R_{b},min}$, $E_{R_{b},max}$) observed in the corresponding APOGEE age bin. 
    It is shown as the black curve in Fig. \ref{apdx:model1_rb-err} for Model1. 
    In other words, each error between $E_{R_{b}, min}$ and $E_{R_{b}, max}$ originally has the same probability of being attributed to a star. 
    We can then sample that error distribution to better represent the observed one. 
    Therefore, the same steps as indicated in section \ref{selection_effects} were applied, the original sample being histogrammed into 0.1 kpc-wide bins. 
    The filled gray histogram in Fig. \ref{apdx:model1_rb-err} shows the resulting error distribution sample.

    \begin{table}
        \caption{Standard deviations of the Gaussian error distributions implemented in the simulations.}
        \label{table:errors}
        \centering
        \begin{tabular}{c c c c c c c}
        \hline\hline
        r & z & v$_{r}$ & v$_{\phi}$ & v$_z$ & age & [Fe/H] \\
        \hline
            0.3 kpc & 0.03 kpc & 2 km/s & 2 km/s & 2 km/s & 10\% & 0.015 dex\\
        \hline
        \end{tabular}
    \end{table}

    \section{Migration over longer time intervals}

\begin{figure*}
        \centering
        {\includegraphics[width=14cm]{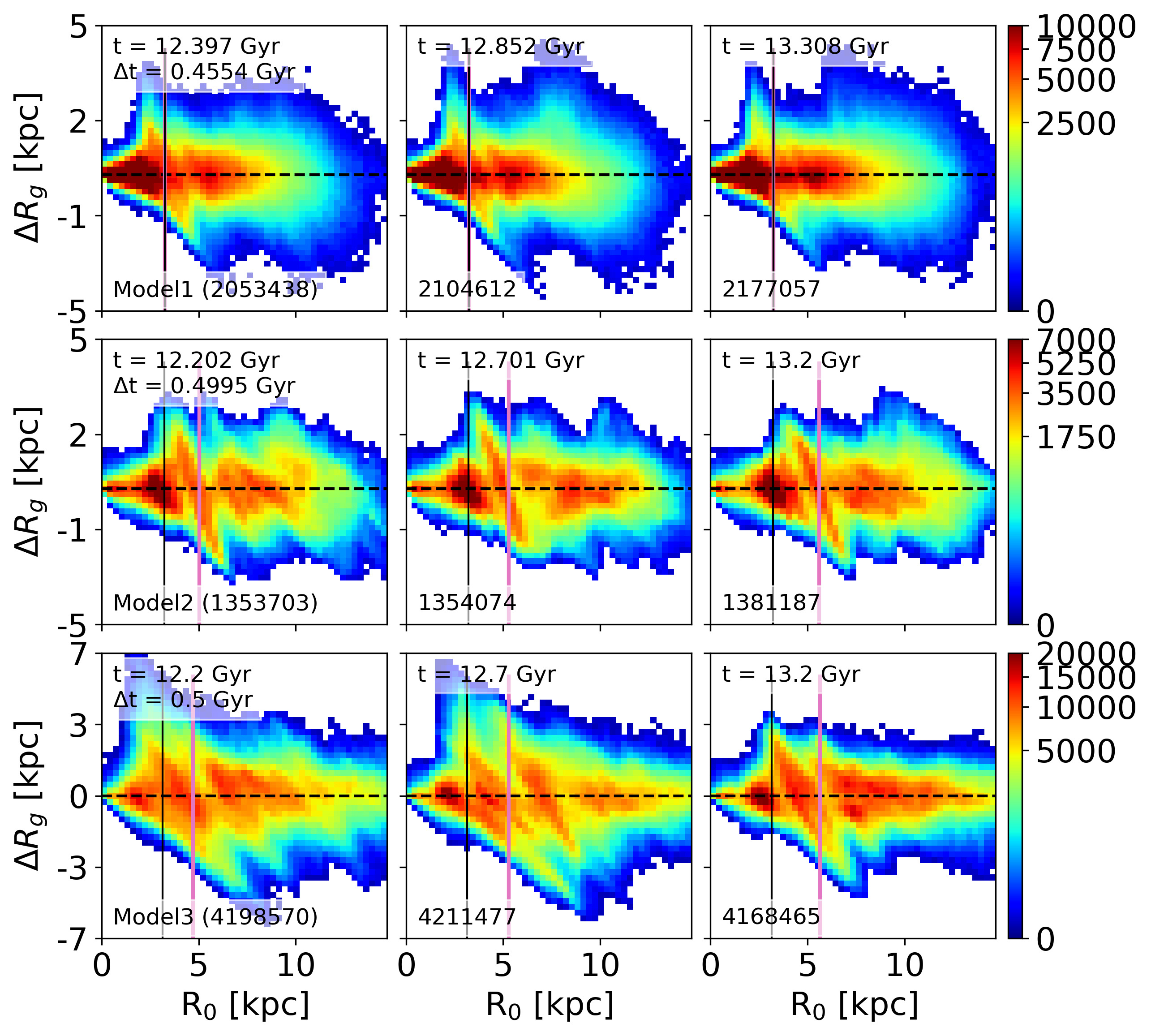}}
        \caption{2D-histograms of the change in guiding radius $\Delta R_{g}$ between two snapshots separated by $\Delta t = 0.5$ Gyr, with respect to the initial guiding radius $R_{0}$. From top to bottom, each row represents stars from Model1, Model2 and Model3. Each column represents a different time in the galaxies' evolution, increasing from left to right. These times are taken randomly, and the time interval is bigger than that used in section \ref{sec:radmigr}. The solid black and pink lines indicate respectively the bar radius and the CR. Strong ridges near these two radii are present in all models. However, the varying amplitude seen in Fig. \ref{fig:drg-r0} are not visible here. Model3 shows decreasing ridge amplitude, most likely as a sign of the galaxy getting hotter with time through numerical heating and absence of cold gas accretion (so less sensitive to perturbations).}
        \label{apdx:drg_r0}
    \end{figure*}

    A usual way to visualize radial migration in simulations is by plotting the change in guiding radius between two snapshots as a function of the initial guiding radius. Changing the time separation between these two snapshots will affect the appearance of these plots. When choosing the right times and the right time interval, the variations in the amplitude of the ridge at bar radius appears just as in Fig. \ref{fig:drg-r0}. However, if the times are not chosen at bar and spiral connection and disconnection and/or if the time interval is too large, these variations caused by the regular bar-spiral overlap will be diluted, as other mechanisms will enter into play and create ridges at other positions. Figure \ref{apdx:drg_r0} shows $\Delta R_{g}$ as a function of $R_{0}$ for a time interval of $\Delta t = 0.5$ Gyr (as opposed to $\Delta t = 0.1$ Gyr for Fig. \ref{fig:drg-r0}). As for Fig. \ref{fig:drg-r0}, each row is for a different model, while each column is for a different time in the evolution of the galaxies. The plane is divided into $0.3$x$3.3$ kpc$^{2}$ bins for Model1 and Model2, and 0.3x4.7 kpc$^{2}$ bins for Model3 and the number of stars in those bins is given by the colorbar, with contribution only from stars present throughout the whole period.
    Strong ridges are seen in all models near the bar's CR, while the ridges at the bar radius are much less pronounced because of the large time interval used here.
    For Model3 (last row), the amplitude of the ridge decreases with time. This most likely results from the disk getting hotter, both numerically and due to the absence of cold gas accretion (in an isolated context), and thus less sensitive to perturbations.

\end{appendix}
\end{document}